\documentclass[12pt]{article}

\usepackage[sorting=none,backend=bibtex]{biblatex}
\usepackage{amsmath}
\usepackage{times}
\usepackage{graphicx}
\usepackage{dcolumn}
\usepackage{bm}
\usepackage{hyperref}
\usepackage{color}
\usepackage{tabularx}
\usepackage{float}
\usepackage{upgreek}
\usepackage{nicefrac} 
\usepackage{enumitem}
\usepackage{placeins}
\usepackage{tabularx}
\usepackage{braket}
\usepackage{acronym}
\usepackage{siunitx}
\usepackage{tikz}
\usepackage{physics}
\usepackage{titling}
\usepackage[capitalise]{cleveref}
\usepackage{lineno}
\usepackage{multirow}
\usepackage{mathtools}
\DeclareUnicodeCharacter{2009}{\,}

\newcommand{\TwoEightSi}{\ensuremath{^{28}\text{Si}}}

\newcommand{\comment}[1]{}

\begin{document}
\title{Waveguide-integrated silicon T centres} 


\author
{A. DeAbreu$^{1,2}$,  C. Bowness$^{1,2}$,  A. Alizadeh$^{1,2}$, C. Chartrand$^{1,2}$,\\ N. A. Brunelle$^{1,2}$,
E. R. MacQuarrie$^{2}$, N. R. Lee-Hone$^{2}$, M. Ruether$^{1,2}$,\\ M. Kazemi$^{1,2}$, A. T. K. Kurkjian$^{2}$, S. Roorda$^{3}$, N. V. Abrosimov$^{4}$,  H.-J. Pohl$^{5}$,\\
 M. L. W. Thewalt$^{1}$, D. B. Higginbottom$^{1,2}$, S. Simmons$^{1,2\dagger}$ \\
\\
\normalsize{$^{1}$Department of Physics, Simon Fraser University,}\\
\normalsize{Burnaby, BC V5A 1S6, Canada}\\
\normalsize{$^{2}$Photonic  Inc., Coquitlam,  BC,  Canada}\\
\normalsize{$^{3}$D\'{e}partement  de physique,  Université  de  Montr\'{e}al,}\\
\normalsize{Montr\'{e}al,  QC H3C 3J7,  Canada}\\
\normalsize{$^{4}$Leibniz-Institut f\"{u}r Kristallz\"{u}chtung, 12489 Berlin, Germany}\\
\normalsize{$^{5}$VITCON Projectconsult GmbH, 07745 Jena, Germany}\\
\\
\normalsize{$^\dagger$Corresponding author. Email: s.simmons@sfu.ca.}
}

\date{}




\baselineskip24pt
\maketitle


\textbf{
The performance of modular, networked quantum technologies will be strongly dependent upon the quality of their quantum light-matter interconnects. Solid-state colour centres, and in particular T centres in silicon, offer competitive technological and commercial advantages as the basis for quantum networking technologies and distributed quantum computing. These newly rediscovered silicon defects offer direct telecommunications-band photonic emission, long-lived electron and nuclear spin qubits, and proven native integration into industry-standard, CMOS-compatible, silicon-on-insulator (SOI) photonic chips at scale. Here we demonstrate further levels of integration by characterizing T centre spin ensembles in single-mode waveguides in SOI. In addition to measuring long spin $T_1$ times, we report on the integrated centres' optical properties. We find that the narrow homogeneous linewidth of these waveguide-integrated emitters is already sufficiently low to predict the future success of remote spin-entangling protocols with only modest cavity Purcell enhancements. We show that further improvements may still be possible by measuring nearly lifetime-limited homogeneous linewidths in isotopically pure bulk crystals. In each case the measured linewidths are more than an order of magnitude lower than previously reported and further support the view that high-performance, large-scale distributed quantum technologies based upon T centres in silicon may be attainable in the near term.
}

Drawing inspiration from classical information processing architectures, \emph{modular} quantum information architectures offer a clear path to large-scale quantum computers and quantum networks. Breaking the constraints of local processors by networking will unlock the full competitive advantage of quantum computing technology. A number of modular quantum architectures have been proposed \cite{Monroe2014,nemoto2014,Lodahl2018} and small-scale demonstrations have been realised \cite{Stephenson2020, Pompili2021a, Gold2021}.

Abstractly, the ideal modular quantum architecture simultaneously offers long-lived, high-fidelity qubits as well as direct, high-quality photonic access to efficiently distribute entanglement across the network. Quantum objects known as spin-photon interfaces offer all of these necessary capabilities in a single quantum element. Examples include trapped ions \cite{Monroe2014,Stephenson2020} and neutral atoms \cite{Leent2022}, but importantly also solid-state systems such as integrated quantum dots~\cite{Ylmaz2010,Greve2012,Gao2012,Arnold2015,Ding2019} 
, rare earth ions \cite{Zhong2018a,Dibos2018,Raha2020,Dibos2022} and colour centre defects within group-IV crystals (diamond \cite{Gruber1997,Hermans2022}, SiC \cite{Falk2013,Wolfowicz2020,Babin2022}, and Si \cite{Yin2013,Morse2017a,Bergeron2020}).

Commercial constraints will affect the pace with which these modular technologies may be deployed in practice. In this context, the ideal quantum architecture interacts with telecommunications-band photons without any transduction \cite{Lauk2020}, and leverages the manufacturing expertise of the semiconductor industry, and in particular silicon, which is the dominant host platform for global integrated photonics and electronics manufacturing. If realised, such an architecture could simultaneously underpin global quantum communications using terrestrial and satellite repeaters~\cite{Borregaard2019, Jennewein2018} as well as distributed modular quantum computers, all leveraging the same shared core quantum element.

Opportunities to realise such an architecture have sparked considerable, but only very recent, interest in telecommunications-band colour centres directly embedded within silicon. There have been three categories of focus: (1) bright emitters without ground state spins, such as the G \cite{Beaufils2018,Prabhu2022,Yang2022,Hollenbach2022}, C \cite{Chartrand2018}, and W \cite{Buckley2017,Baron2022} centres, (2) dim yet high-efficiency emitters with optical access to spins such as erbium \cite{Yin2013,Yang2022,Berkman2021,Weiss2021,Weiss2021}, and (3) the T centre, which offers relatively bright photon emission as well as direct access to long-lived spins. The T centre was first assessed and proposed as a competitive candidate quantum spin-photon interface in 2020 \cite{Bergeron2020}. Integration tests quickly followed, establishing that T centre qubits could be created at high densities~\cite{MacQuarrie2021} and individually addressed at scale in industry-standard silicon-on-insulator (SOI) \cite{Higginbottom2022}.

Although the T centres in these demonstrations possessed many encouraging features, it was unclear if their optical properties were sufficient to support optically-generated multi-qubit entanglement. Spin-photon interfaces may be entangled by emission~\cite{Barrett2005} or reflection-based \cite{Waks2006,Kerckhoff2009} protocols. Emission-based protocols require that the ratio of homogeneous and lifetime-limited linewidths is close to one~\cite{Kambs2018, Bylander2003}. In general this `transform limit' may be reached by utilizing optical resonators to Purcell enhance the emission and increase the lifetime-limited linewidth accordingly~\cite{Martini1987,Wein2018,Dibos2018}. Similarly, reflection-based protocols require that homogeneous linewidth is small compared to the resonator-emitter coupling. In both cases the homogeneous linewidth critically determines the optical resonator quality required for high fidelity entanglement, and for T centres the relevant homogeneous linewidth remained unknown.  

In this work we study the optical and spin properties of T centres in integrated silicon photonics. We fabricate single-mode waveguides containing ensembles of T centres and measure their key optical properties. We find remarkably sharp homogeneous linewidths that will support remote entanglement generation with modest Purcell enhancements. We compare these linewidths with new measurements in bulk \TwoEightSi{} and report near-transform-limited optical transitions. Finally, we place our integrated sample in a magnetic field and use hole burning to initialize and readout the spin states and report competitive spin $T_1$ lifetimes.

\begin{figure}
    \centering
    \includegraphics[width=17.2cm]{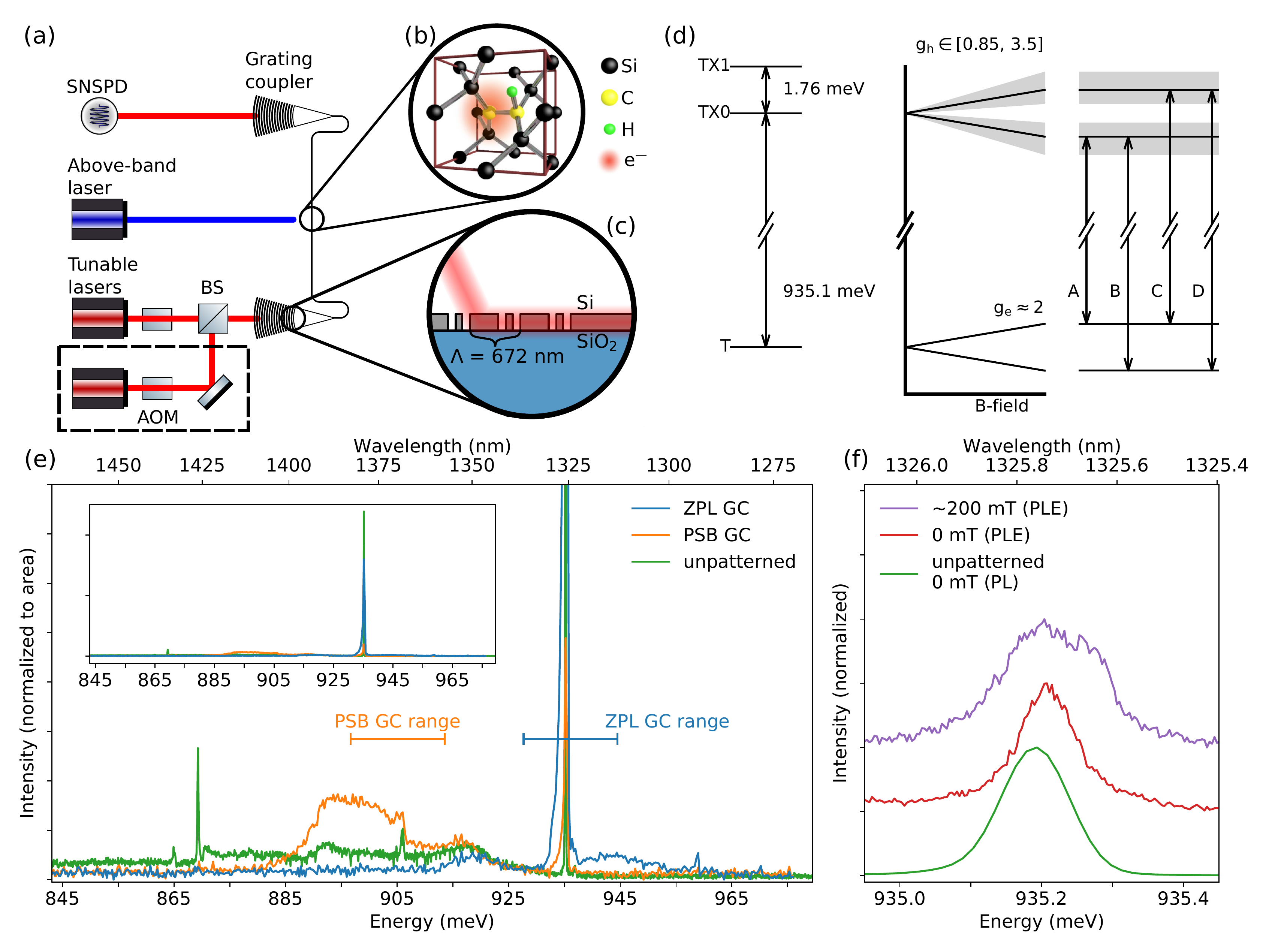}
    \caption{\textbf{T centre waveguide integration and optical spectra.} \textbf{(a)} Experimental setup for measuring T centre ensembles within integrated waveguide devices (not to scale) illustrating the two excitation pathways: resonantly coupled by a GC through the waveguide with an optional second resonant laser and above-bandgap light from above. \textbf{(b)} The proposed chemical structure of the T centre. \textbf{(c)} A side-on schematic of the subwavelength grating coupler. \textbf{(d)} Energy level diagram of the T centre both at zero field and with a magnetic field. \textbf{e)} PL spectra from CW above-bandgap excitation measured through the two GC ports in addition to a PL spectrum from an identical unpatterned sample. The nominal wavelength ranges of the zero-phonon line and phonon sideband grating couplers are marked by horizontal bars. \textbf{(f)} PLE spectra of the TX$_0$ ZPL line for the same device both in a magnetic field and at zero field compared to the PL of the unpatterned material.}
    \label{fig:device}
\end{figure}

\textbf{The T centre.}
The T centre is a multi-component colour center in silicon, possessing a bound exciton optical transition at $1326$~nm ($935$~meV), in the telecommunications O-band~\cite{Bergeron2020}. Known for many years as a defect in the class of radiation damage centres, many of the T centre's properties had been known previous to its rediscovery as a qubit candidate~\cite{Safonov1996b, Minaev1981a}. The proposed chemical structure~\cite{Safonov1996b}, shown in \cref{fig:device}(b), comprises two inequivalent carbon atoms and a hydrogen atom, and possesses an unpaired electron in the ground state with a isotropic Land\'{e} g-factor~\cite{Bergeron2020}. In the excitonic ground state the electron of the bound exciton forms a singlet with this ground state electron. The expected fourfold degeneracy of the spin-3/2 hole is lifted by the reduced symmetry of the T centre, resulting in a spin-1/2 doublet ground state for the bound exciton, TX$_0$, and a higher energy doublet, TX$_1$, $1.76$~meV above TX$_0$.

The TX$_0$ to ground state transition was determined to have a lifetime of $0.94\pm0.01$~$\upmu$s and a Debye-Waller factor of $0.23\pm0.01$ \cite{Bergeron2020}. The radiative efficiency is not precisely known, with theoretical estimates in the range $0.19$--$0.72$ \cite{Dhaliah2022} and an experimental lower bound of $0.03$ \cite{Higginbottom2022}. Thermal activation from TX$_0$ to TX$_1$ was found to freeze out below $\sim2$~K resulting in low temperature inhomogeneous linewidth of $6.5\pm0.2$~GHz for silicon with a natural isotopic abundance and $33\pm2$~MHz for isotopically purified \TwoEightSi{} \cite{Bergeron2020}.

Spectroscopy at magnetic fields revealed 12 inequivalent sets of 24 possible T centre orientations. The ground state electron Land\'{e} g-factor was found to be isotropic with a value of $2.005\pm0.008$ whereas the hole g-factor was found to be very anisotropic, ranging between $0.85$--$3.5$, depending on the magnetic field axis relative to the axis of the orientational subset \cite{Bergeron2020, MacQuarrie2021}. Furthermore, the ground state level structure was determined to feature an anisotropic hyperfine coupling between the electron and the spin-1/2 hydrogen nucleus with an effective hyperfine splitting $<5$~MHz.

\textbf{T centre waveguide devices.} 
We generate T centres in a silicon-on-insulator (SOI) wafer and pattern single-mode waveguide devices (see Methods). These devices are optically addressed by an array of single-mode fibres positioned above the plane of the photonic chip, which is optionally mounted with a removable permanent magnet and cryogenically cooled to either $4.3$~K or $1.2$~K (Methods). Each device is a $360$~$\upmu$m long single-mode strip waveguide, terminated by two sub-wavelength grating couplers (GCs)~\cite{Wang2015}, shown schematically in \cref{fig:device}(a). The two GCs are designed for different wavelengths, one centred on the T centre zero-phonon line (ZPL) and one red-shifted GC that covers a portion of the phonon sideband (PSB) spectrum. This configuration allows for resonant excitation of the TX$_0$ ZPL through one GC port and detection of phonon sideband fluorescence through the other. The sample can be resonantly excited with one or two, independently pulsed, tunable lasers. Emission is collected from the PSB GC and directed to an external superconducting nanowire single photon detector (SNSPD). An additional fibre port directly above the waveguide centre can be used for direct, above-bandgap excitation of the waveguide itself, with fluorescence collected from either the PSB or ZPL GC ports.

\textbf{PL and PLE.} We first demonstrate T centres within the waveguide by photoluminescence (PL) spectroscopy. Above-bandgap excitation through the middle port generates excitons that bind to centres and subsequently luminesce upon recombination. We measure the luminescence separately from the two GC ports and record spectra with an ultra low-noise optical spectrometer. \Cref{fig:device}(e) shows these two PL spectra compared to a reference spectrum recorded from an identical unpatterned SOI chip. The ZPL GC spectrum is dominated by a bright TX$_0$ ZPL, and the PSB GC spectrum exhibits the distinctive TX$_0$ local vibrational mode (LVM) replicas $L_1$,$L_2$ and $L_3$ at $906$, $869$ and $865$~meV respectively~\cite{Bergeron2020, Safonov1996b}.

We measure the T waveguide device at higher resolution by photoluminescence excitation (PLE) spectroscopy. \Cref{fig:device}(f) compares the waveguide T PLE spectrum with the unpatterned material spectrum from \cref{fig:device}(e) in the vicinity of the TX$_0$ ZPL. The resulting zero-field PLE spectrum is typical of an  inhomogeneous T centre ensemble. Concentration estimates on similar material \cite{Higginbottom2022} suggest that each waveguide device contains upwards of $600$ centres. Resonant excitation through the ZPL GC illuminates the entire waveguide with a single linearly polarized optical mode. The inhomogeneous distribution of the T centre ensemble within this waveguide is slightly narrower than the unpatterned material, $22.7\pm0.3$ and $27.3\pm0.2$~GHz respectively, indicating that the TX$_0$ energy is insensitive to waveguide fabrication. Fabricating devices may even release strain from the SOI chip \cite{Prabhu2022}.

\begin{figure}
    \centering
    \includegraphics[width=8.6 cm]{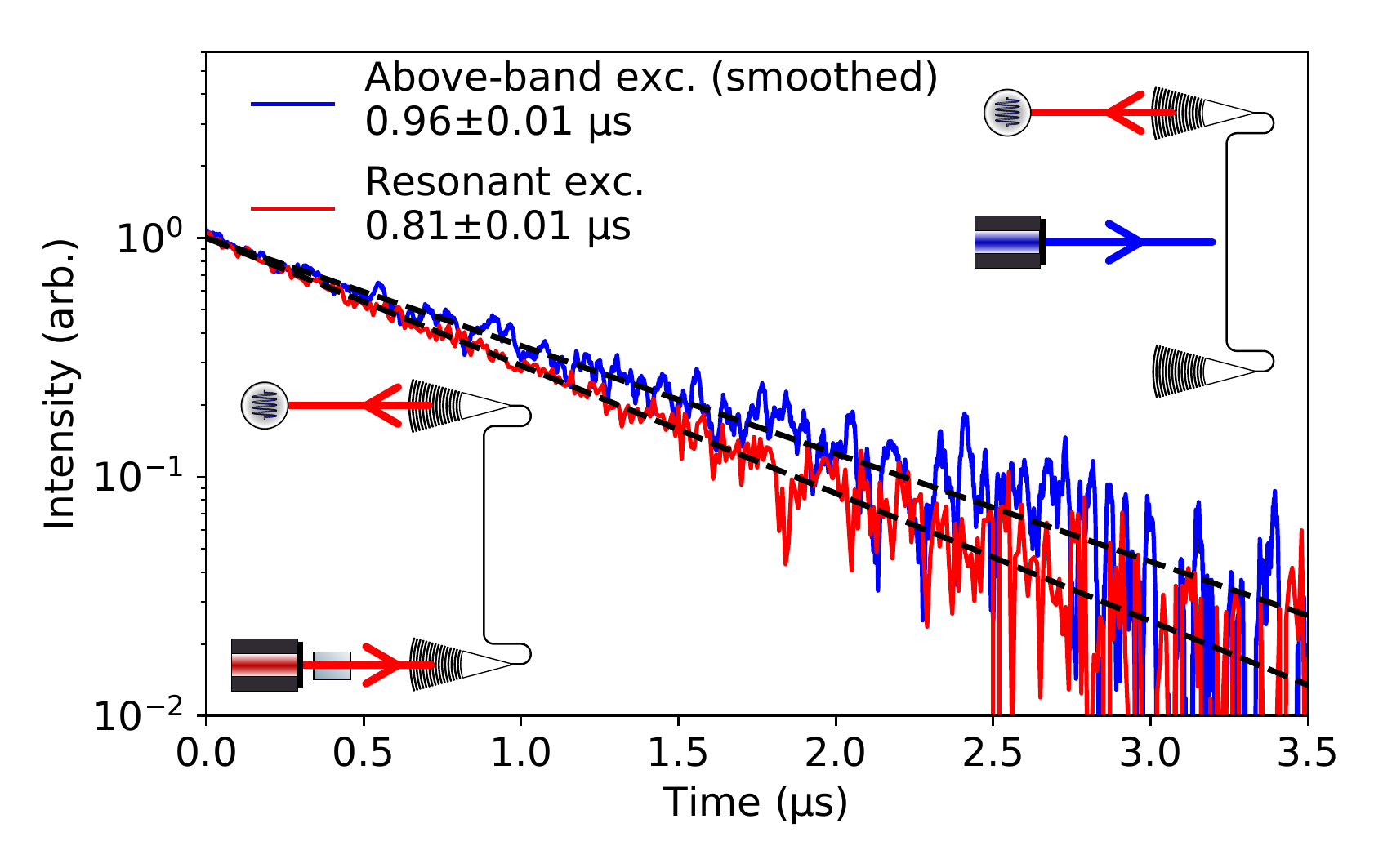}
    \caption{\textbf{Lifetime measurements.} Luminescence lifetime from the two excitation pathways: pulsed above-bandgap light coupled from above (blue, upper right inset) and pulsed resonant light coupled through the waveguide (red, lower left inset). Both curves are fit with an exponential decay.
    }
    \label{fig:lifetime}
\end{figure}

\textbf{Lifetime measurements.} We measure the average excited state lifetime of this waveguide-coupled ensemble and find that it is not dramatically different from the bulk crystal, but the precise value depends on the excitation method. First, we measure the lifetime with above-bandgap excitation through the middle fibre using a $965$~nm pulsed laser. The transient fluorescence, shown in \cref{fig:lifetime}, decays exponentially with a lifetime of $0.96\pm0.01$~$\upmu$s, consistent with the equivalent bulk above-bandgap excitation lifetime of $0.94\pm0.01$~$\upmu$s \cite{Bergeron2020}. We conclude that there is no evidence of Purcell enhanced or suppressed emission within this measurement uncertainty (simulated waveguide Purcell factors are included as SI). 

When exciting resonantly through the waveguide we measure a lifetime of $0.81\pm0.01$~$\upmu$s, 16$\%$ lower than the above-bandgap excitation value. The excited state lifetime shows no dependence on the resonant excitation power (SI \cref{fig:sm_lifetime}) indicating that stimulated emission does not account for the difference. It may instead be due to the average free exciton capture time or superradiant enhancement \cite{Goban2015, Lukin2020superradiance}.

\textbf{Spectral hole burning.} An emitter's homogeneous linewidth is a critical metric for entangling and networking spin-photon qubits \cite{Kalb2018DephasingNetworks}. An ideal emitter will have a lifetime-limited linewidth, however this deteriorates with dephasing and spectral diffusion due to environmental fluctuations. In Ref.~\cite{MacQuarrie2021} the \emph{long-time} homogeneous linewidth of unpatterned SOI with the same T centre recipe was determined to be $1.3\pm0.3$~GHz, limited by `slow' spectral diffusion (SD) on timescales much longer than the excited state lifetime. Similar long-time homogeneous linewidth values for single-T centres were measured by Ref.~\cite{Higginbottom2022}. Such slow SD can be overcome with a feedback mechanism that tunes wandering emitters back into mutual resonance~\cite{Acosta2012, Bernien2012}. Dephasing and `fast' SD (on the timescale of the emission lifetime) on the other hand, are an intrinsic limit to emitter performance~\cite{Kalb2018DephasingNetworks, Kambs2018}. Here we report the first \emph{instantaneous} homogeneous linewidth measurement of T centres. We compare the instantaneous linewidth in both waveguide devices and bulk isotopically-enriched \TwoEightSi{} by spectral hole burning and find linewidths that are dramatically lower than the published long-time linewidths and very promising for T centre spin-photon networks.

A pump laser is tuned to the peak of the inhomogeneous linewidth and a second probe laser is swept about this wavelength. When the lasers are detuned further than a homogeneous linewidth, $\Delta\omega_{\rm hom}$, the lasers address independent sub-ensembles of T centres and produce a fluorescence signal that is simply the sum of the emission from each sub-ensemble. When the detuning is small and the two lasers are within one homogeneous linewidth, the lasers address the same sub-ensemble and the signal is reduced due to saturation. Scanning the probe laser produces a spectrum with a `spectral hole' at the wavelength of the pump laser, and linewidth, $\Delta\omega_{\rm hole}$, given by:
\begin{equation}
    \Delta\omega_{\rm hole} = \Delta\omega_{\rm hom}\left( 1 + \sqrt{1 + \left(P_{\rm probe}+P_{\rm pump}\right)/P_{\rm sat}}\right)
    \label{eq:hole_burning}
\end{equation}
where $P_{\rm probe}$, $P_{\rm pump}$, and $P_{\rm sat}$ are the probe, pump, and saturation powers respectively \cite{Siegman1986}. In general $\Delta\omega_{\rm hom} < \frac{1}{2}\Delta\omega_{\rm hole}$.

\begin{figure}
    \centering
    \includegraphics[width=8.6cm]{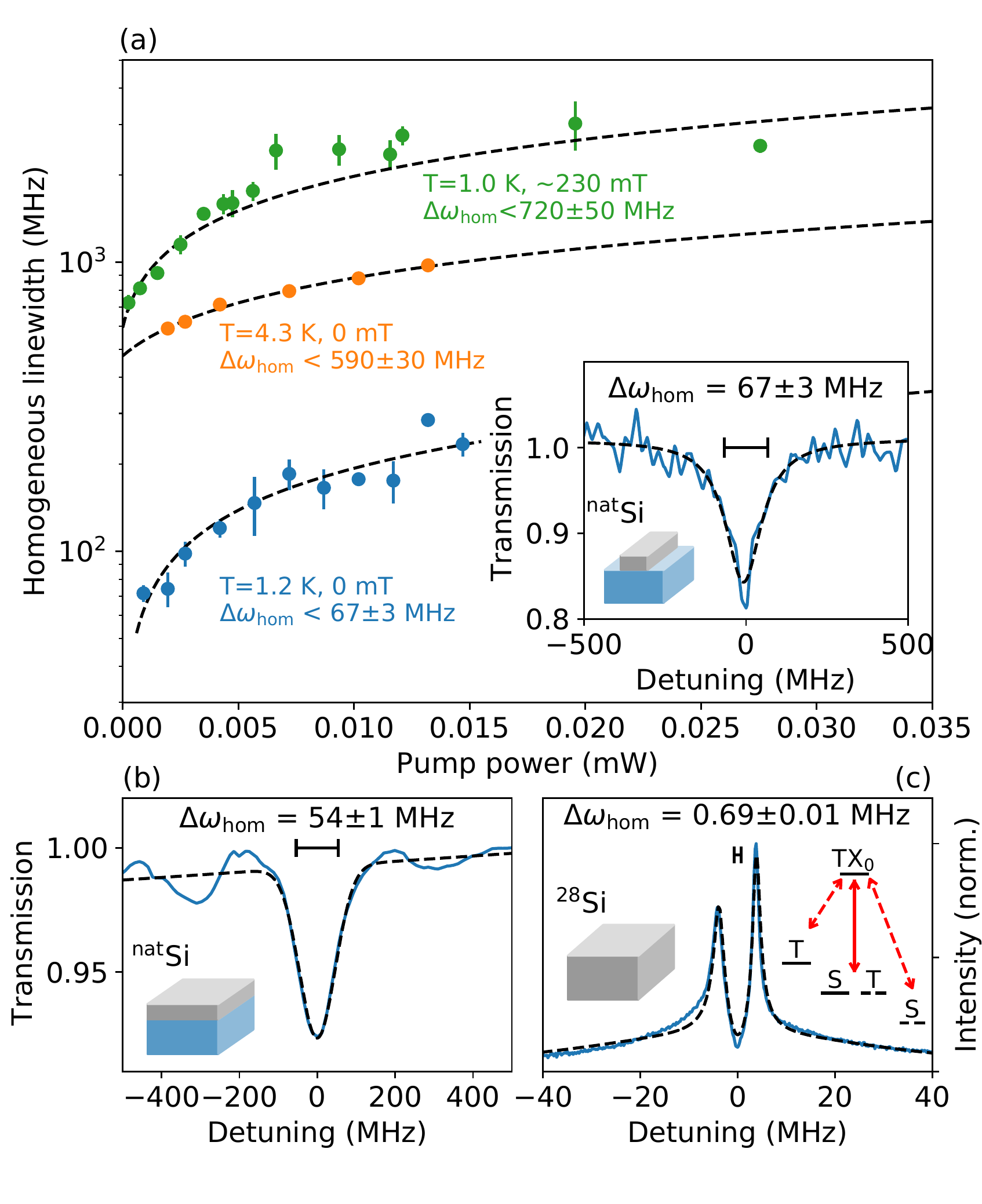}
    \caption{\textbf{Instantaneous homogeneous linewidth via hole burning.} (a) Spectral hole burning in a waveguide device under various temperature and field conditions, all fit with optical pump power dependence from Eq.~\ref{eq:hole_burning}. (Inset) Lowest pump power spectral hole measured at $1.2$~K, zero-field. (b) Hole burning on unpatterned SOI shows only a small linewidth increase from device patterning. (c) Hole burning on bulk \TwoEightSi{}, fit with a three level rate equation. (Inset) Triplet-singlet energy level for two sub-ensembles showing the pump-probe detunings for the anti-holes.
    }
    \label{fig:hole_burning}
\end{figure}

 We perform spectral hole burning on waveguide devices at $1.2$~K, at zero magnetic field, at a range of pump powers, and measure an instantaneous linewidth that power broadens according to Eq.~\ref{eq:hole_burning}, shown in \cref{fig:hole_burning}(a). The spectral hole we measure at the lowest pump power, inset to \cref{fig:hole_burning}(a), corresponds to an instantaneous linewidth of only $67\pm3$~MHz---an order of magnitude lower than the long-time single-T centre device linewidths measured in Ref.~\cite{Higginbottom2022}. We note that this measurement does not attain the low power limit, as such it is an upper bound for the homogeneous linewidth. Extrapolating the fitted linewidth model to the low-power limit, we infer that the true instantaneous linewidth could be as low as $11\pm11$~MHz.

 We can illustrate the impact of additional dephasing by measuring waveguide devices at an elevated temperature. At $1.2$~K, thermal activation to TX$_1$ is frozen out, but at $4.3$~K thermal excitation between the two excited state levels dephases the optical transition TX$_0$ \cite{Bergeron2020}. We observe a minimum instantaneous linewidth of $590\pm30$~MHz at $4.3$~K and once again measure optical power broadening according to Eq.~\ref{eq:hole_burning}. The inferred low-power limit is $470\pm30$~MHz, higher than the $290$~MHz expected at this temperature from the model in Ref.~\cite{Bergeron2020}. PL spectra confirm that the TX$_0$--TX$_1$ level splitting is unchanged compared to bulk ensembles. The difference may instead be due to the altered phononic density of states within the waveguide.

The low-temperature linewidth is a significant improvement over the performance used to estimate requirements for a T centre quantum optical network in Refs.~\cite{MacQuarrie2021,Higginbottom2022}, and very promising for an integrated luminescence center with little material or fabrication process optimization yet done. The primary reason these hole burning results produce narrower linewidths than those measured in Refs.~\cite{MacQuarrie2021,Higginbottom2022} is the measurement timescale. The burnt hole heals during the centre's excitation and decay time, approximately the excited state lifetime ($<1$~$\upmu$s). Hole burning provides a measure of the instantaneous homogeneous linewidth depending only on dephasing and spectral diffusion from environmental fluctuations happening within this short window. The relevant timescale for Refs.~\cite{MacQuarrie2021} and \cite{Higginbottom2022} are, respectively, the electron spin $T_1$ lifetime and the measurement integration time---both much longer than $1~\upmu$s. T centre material and fabrication studies will likely achieve further reductions by removing sources of spectral diffusion, as has been seen in other materials~\cite{Anderson2019, vandam2019}.
 
 We compare instantaneous linewidths before and after waveguide fabrication by performing zero-field, low-temperature hole burning on unpatterned SOI (see SI for details). This hole burning spectrum, shown in \cref{fig:hole_burning}(b), indicates a $54\pm1$~MHz instantaneous linewidth. We conclude that neither the electron beam lithography and etching procedure to produce waveguide devices, nor the presence of new, rougher, etched silicon-air interfaces results in greater fast spectral diffusion. The resilience of T centres to integrated device environments bodes well for future work with T centres and photonic interfaces, such as the integration with nanophotonic cavities. 

For comparison, we measured the hole burning spectrum of a \TwoEightSi{} sample to obtain the linewidth in an isotopically and chemically pure environment. Strikingly, this instantaneous linewidth measurement approaches the T centre's fundamental lifetime-limited linewidth. An example \TwoEightSi{} hole burning spectrum is shown in \cref{fig:hole_burning}(c). It features two sharp peaks or `anti-holes' with only a faint indication of a hole. This indicates that we are able to hyperpolarize into each of two long-lived ground states. We are optically resolving, for the first time, the zero field hyperfine splitting of the ground state \cite{Bergeron2020} (illustrated in the inset of \cref{fig:hole_burning}(c)). 

A three level rate equation model was created to determine the true instantaneous linewidth from this steady-state spectrum. The full details of this model are presented in the SI which, in addition to the instantaneous linewidth, had the zero-field hyperfine splitting and the relative degeneracies of the ground states as free parameters. This model was used to simultaneously fit five hole burning spectra at different pump/probe power ratios, resulting in a triplet-singlet zero field splitting of $3.85\pm0.01$~MHz, with the triplet higher in energy than the singlet. Finally, the fit revealed a instantaneous linewidth of $0.69\pm0.01$~MHz---just $4$ times larger than the lifetime-limited value of $170$~kHz. Only very modest lifetime reduction would be needed to obtain transform-limited emission from this \TwoEightSi{} base material. These measurements indicate that the T centre's instantaneous linewidth can be made much lower than previously known \cite{Bergeron2020}, and further improvements to the T centre implant recipe and SOI material may yield a substantial linewidth reduction.

\begin{figure}
    \centering
    \includegraphics[width=17.2 cm]{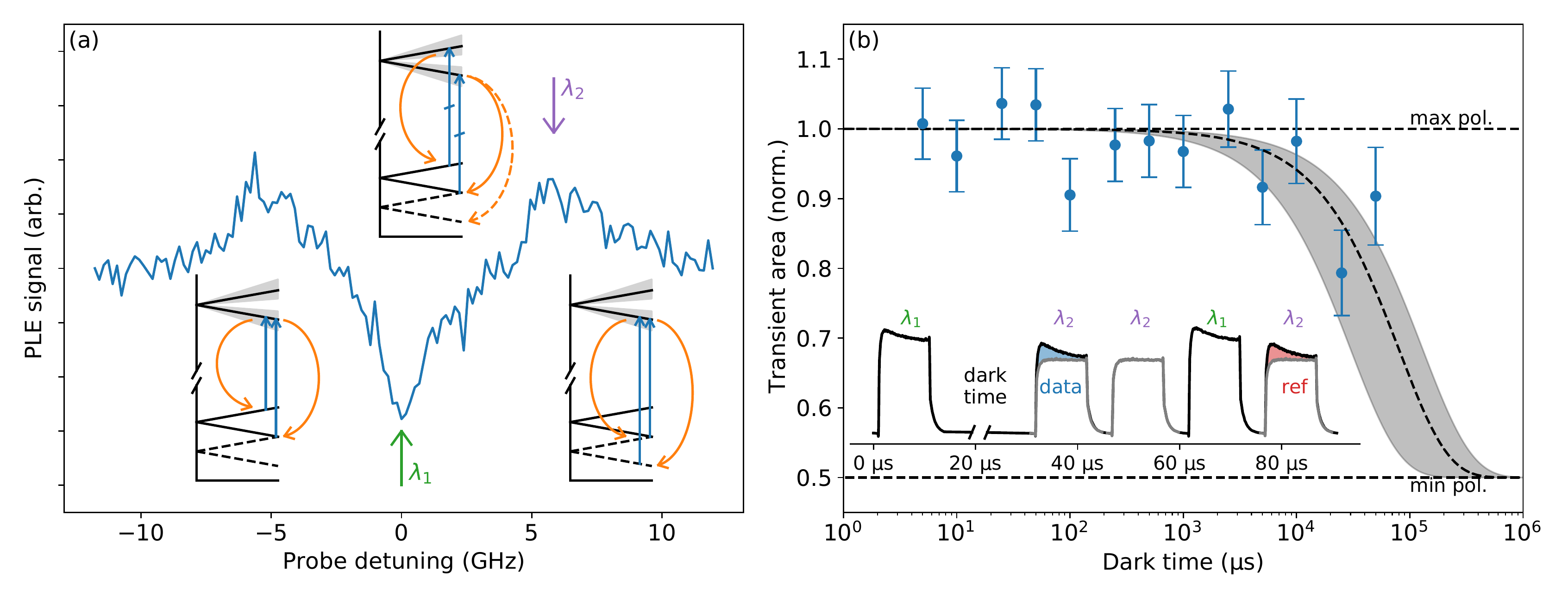}
    \caption{\textbf{$\mathbf{T_1}$ measurement via hole burning.} (a) Hole burning spectrum of the sample at field showing the saturation hole at zero detuning and the anti-holes at detunings equal to the ground state splitting. (Insets) Energy level diagrams schematically illustrating the pump-probe conditions for saturation and anti-holes for two sub-ensembles of the inhomogeneously broadened line (solid and dashed). At zero detuning the horizontal dashes indicate equal energies. (b) Spin relaxation measured over time bounds the population lifetime $T_1 > 80\pm50$~ms. (Inset) The pulse sequence of the two lasers tuned to $\lambda_{1,2}$ as shown in (a) along with the measured signal from each pulse showing the hyperpolarization transient. Spin relaxation is determined by the ratio of data (blue) and reference (red) transient areas.
    }
    \label{fig:figure_4}
\end{figure}

\textbf{Spin manipulation.} In order to optically control the ensemble's electron spin state we may apply a magnetic field and split TX$_0$ into four spin-selective transitions A--D as per \cref{fig:device}(d). When mounted on the permanent magnet we expect a $\sim200$~mT magnetic field at the device position, oriented normal to the SOI sample along the $\langle100\rangle$ axis. In this configuration the inhomogeneous spectrum broadens to $39.6\pm0.5$~GHz (\cref{fig:device}(f)), which is not accounted for by the magnetic field alone. Clamping the sample over the neodymium magnet may strain the sample and shift TX$_0$ across the waveguide ensemble~\cite{Ciechanowska1984, stoneham1969}. At this field the ground state is split by $\Delta_\mathrm{g}=6$~GHz and the excited state is split by either $\Delta^1_\mathrm{e}=3$~GHz or $\Delta^2_\mathrm{e}=8$~GHz depending on centre orientation ($g^{1,2}_\mathrm{h}=0.91,2.55$)~\cite{MacQuarrie2021}. We therefore expect well-resolved spin-selective transitions for a subset of centres in the device, allowing us to hyperpolarize the ground state electron just as we did in bulk \TwoEightSi{} at zero-field. 

We confirm the ability to hyperpolarize a subset of the sample by observing anti-holes in the hole burning spectrum wherein both of the ground states are being addressed by either the pump or probe lasers. \Cref{fig:figure_4}(a) shows an example spectrum with anti-holes separated by $\sim12$~GHz. We also observe a spectral hole, indicating a subset of centres with sufficiently large B--C overlap to produce a saturation hole. The power dependence (see \cref{fig:hole_burning}(a)) is well modelled by Eq.~\ref{eq:hole_burning} and we observe a minimum instantaneous linewidth of $720\pm50$~MHz. An increase in instantaneous linewidth is to be expected; measuring the hole at field preferentially selects centres with instantaneous linewidth on the order of the B--C splitting, $\Delta_\mathrm{BC}^{1,2} = \Delta_\mathrm{e}^{1,2} - \Delta_\mathrm{g} = \{2,~3\}$~GHz, as these centres contribute disproportionately to the single-laser fluorescence. It is also possible that the observed sample strain increases spectral diffusion.

In general we expect anti-holes at pump-probe detunings of $\pm\Delta_\mathrm{g}, \pm\Delta_\mathrm{BC}$ and $\pm\Delta_\mathrm{AD} = \pm (\Delta_\mathrm{g} + \Delta_\mathrm{e})$~\cite{Nilsson2004}. The observed $\pm6$~GHz anti-holes splitting is attributed to $\Delta_\mathrm{e}$. We confirm this assessment with a zero-parameter fit of a four level rate equation model included as SI. With this assignment in mind we schematically illustrate the formation of the hole and anti-holes in the insets of \cref{fig:figure_4}(a). A saturation hole is present when both pump and probe are driving the overlap of the B and C transitions. The anti-holes are symmetric about the hole as inhomogeneous broadening will lead to two bright subsets illustrated with solid and dashed ground state levels.

This hole-burning spectrum indicates that a subset of T centre spins can be initialized optically. Additionally, a pulsed pump laser can be used to measure the spin population via the luminescence transient as the spins hyperpolarize from bright to dark states. This is sufficient to measure a $T_1$ lifetime for this subset \cite{Higginbottom2022}. We apply two lasers pulsing at wavelengths $\lambda_1$ and $\lambda_2$ (as indicated in Fig.~\ref{fig:figure_4}) separated by the ground state splitting $\Delta_\mathrm{e}$. The pulse scheme and luminescence transients are shown inset to \cref{fig:figure_4}(b). The system is initialized into an out-of-equilibrium spin state in which a sub-ensemble is hyperpolarized by $\lambda_1$. After a variable dark time the spin state is read out with two pulses at wavelength $\lambda_2$. The first `readout' pulse will pump between the two electron spin states of the sub-ensemble, leading to a transient luminescence signal. Subsequently probing the system with the same wavelength ($\lambda_2$) yields no further hyperpolarization or accompanying transient, but rather only a constant background luminescence signal from the unresolved sub-ensemble. The difference between the integrated luminescence during these pulses is proportional to the spin population in the addressed state. This readout transient (blue) is labeled `data'. Afterwards, an additional round of initialization and readout is performed without any dark time to give a reference transient (red), labeled `ref'. 

The data transient is compared to the reference to get a polarization ratio that drops from 1 (maximally polarized) to 0.5 (no polarization) in thermal equilibrium. The resulting dependence of the polarization ratio versus dark time is shown in \cref{fig:figure_4}. We observe a polarization decay from which we fit a $T_1$ spin lifetime of $80\pm30$~ms. This effective $T_1$ is a lower bound to the true spin lifetime. Due to finite optical pulse extinction, leaked laser light during the dark time will slowly drive the spins to a mixed state faster than the true relaxation time. 

We have therefore demonstrated all-optical initialization and readout of electron spins in a waveguide-coupled T centre ensemble. Although the measured bound $T_1 > 50$~ms is short compared to the bulk relaxation time, $T_1 > 16$~s \cite{Bergeron2020}, it is sufficiently long to measure the electron spin coherence by combining the same optical initialization and readout scheme with microwave control of the T centre ground state.

\textbf{Discussion.} We have integrated the silicon T centre with monolithic photonic waveguides and measured the homogeneous and ensemble properties. Such T centre devices can be efficiently networked on chip with integrated photonic cavities, switches, and detectors \cite{Akhlaghi2015}, as well as efficiently fibre coupled for remote networks. Integrating the T centre with nanophotonics fabricated by a standard, unoptimized commercial CMOS process shows no significant degradation of optical properties compared to the unpatterned chip material. Most promisingly, we measure instantaneous homogeneous linewidths an order of magnitude better than spectral diffusion linewidths previously reported. New instantaneous homogeneous linewidth measurements performed by hole-burning on a bulk \TwoEightSi{} sample show that a nearly transform-limited linewidth is achievable.

With these instantaneous homogeneous linewidth values we can, for the first time, predict the indistinguishability of networked T centre spin-photon interfaces. With appropriate feedback the slow spectral diffusion can be made negligible and luminescence from two or more emitters with zero relative detuning can be made to interfere ~\cite{Acosta2012, Bernien2012}. In this limit interference visibility is determined only by the emitter lifetimes and the emission linewidth~\cite{Kambs2018}. We determine the interference visibility for T centres in $^{\rm nat}$Si waveguides and bulk $^{\rm 28}$Si based on our measurements (SI, Tab.~\ref{tab:visibilities}). We also tabulate what Purcell factor, $F_{P}$, and cavity quality, $Q$, are necessary to entangle two T centre spins beyond the Bell threshold \cite{Hensen2015}, assuming unit radiative efficiency, and find it is achievable with demonstrated photonics in every case. For waveguide-integrated T centres, based on the lowest measured linewidth (or inferred low-power limit), only $F_{P} \approx 2,200$ ($350$), achievable in a nanophotonic cavity of $Q~\approx14,000$ ($2,300$) is required. Such quality factors are routine in silicon nanophotonics~\cite{Ashida2017,Deotare2009}. Centres from the \TwoEightSi{} ensemble could achieve a $25\%$ visibility without any enhancement, and require only $F_{P}=13$ to reach the Bell violation threshold.

 We showcased the ability to optically initialize and readout electron spins within the waveguide T centre ensemble. We used an all-optical pulse sequence to measure spin relaxation and find a competitive bound on $T_1$ lifetime. This is a crucial first step towards full spin control of waveguide-device integrated T centres, requiring only inclusion of microwave lines for driving the spins directly. These new measurements make the T centre an encouraging spin-photon interface for on-chip networks, ready to be deployed in a silicon-based modular quantum information architecture. 


\textbf{Acknowledgments}
The $^{28}$Si samples used in this study were prepared from Avo28 crystal produced by the International Avogadro Coordination Project (2004–2011) in cooperation among the BIPM, the INRIM (Italy), the IRMM (European Union), the NMIA (Australia), the NMIJ (Japan), the NPL (United Kingdom), and the PTB (Germany).

The authors would like to thank Dr. Chloe Clear for beneficial discussion during the drafting of this manuscript.

\textbf{Funding:} This work made use of the 4D LABS and Silicon Quantum Leap facilities supported by the Canada Foundation for Innovation (CFI), the British Columbia Knowledge Development Fund (BCKDF), Western Economic Diversification Canada (WD) and Simon Fraser University (SFU). This work was supported by the Canada Research Chairs program (CRC), the New Frontiers in Research Fund: Exploration (NFRF-E), the Canadian Institute for Advanced Research (CIFAR) Quantum Information Science program and Catalyst Fund, Le Fonds de recherche du Qu\'{e}bec – Nature et technologies (FRQNT), and the Natural Sciences and Engineering Research Council of Canada (NSERC). D.B.H. is supported by an NSERC Banting Fellowship.

\textbf{Author Contributions:} A.D., D.B.H., and S.S designed the experiments and wrote this manuscript. A.D., C.B., and D.B.H., built the experimental apparatus and measured waveguide T centre spectra. A.D. measured optical and  spin lifetimes. A.D. and A.A. measured  hole burning in waveguides. A.A. and C.C. measured hole burning in bulk SOI. A.A., N.B., and A.D. measured hole burning in \TwoEightSi{}. A.D. performed the Lumerical simulations, developed the rate equation model, and performed the data analysis. E.R.M., D.B.H, C.C., and S.R. developed the SOI samples. C.B. and D.B.H. designed the photonic devices and chip. N.V.A. and H.-J.P. developed the \TwoEightSi{} sample. N.R.L.-H., M.K., and M.R. assisted in the experimental design. A.D., C.B., D.B.H, A.A., C.C., N.A.B., E.R.M., N.R.L.-H., M.R., M.K., and A.T.K.K contributed  to the code development. M.L.W.T. advised and assisted with sample creation, experimental design and analysis. All authors participated in manuscript revision.

\textbf{Supplementary Information} is available for this paper. 

\textbf{Data and materials availability:} Data is available on request. Correspondence and requests for materials should be addressed to Stephanie Simmons.

\printbibliography

\section*{Methods}

\paragraph*{Photonic waveguide chip.} All waveguide measurements are performed on a sample created by generating T centres in an unpatterned, commercially available Czochralski silicon-on-insulator (SOI) wafer with a $220$~nm device layer (P-type, $50$--$100$~Ohm-cm), and a $3$~$\upmu$m buried oxide layer atop a $700$~$\upmu$m silicon handle layer. T centres were generated by implanting carbon and hydrogen sequentially. Carbon-12 implantation was performed by Innovion, with a dose of $7 \times 10^{13}~\rm{cm}^{-2}$ at an energy of 38 keV followed by a rapid thermal anneal at 1000~°C for 20~s in an argon background performed by Marvell Nanofab. The sample was then returned to Innovion for hydrogen implantation at a dose of $7 \times 10^{13}~\rm{cm}^{-2}$ at an energy of 9 keV.  After implantation the 8" diameter wafer was diced into 2~cm x 2~cm chips using a rough cut manual cleave followed by a precision deep etch dicing process. Dicing was followed by an anneal in 100~$^\circ$C de-ionized water for 1 hour and a rapid thermal anneal for 3 minutes at 405~°C in a nitrogen background. The photonic devices were fabricated via electron beam lithography and an anisotropic plasma etching process performed by Applied Nanotools (ANT).

\paragraph*{Unpatterned SOI} The unpatterned SOI hole burning measurements are performed on an unpatterned, commercially available Czochralski silicon-on-insulator wafer with a 220 nm device layer (P-type, 50-100 Ohm-cm). A dose of $7 \times 10^{13}~\rm{cm}^{-2}$ carbon-13 was implanted at an energy of 38 keV by Cutting Edge Ions followed by a rapid thermal anneal at 1000~°C in an argon background performed by Washington Nanofabrication Facilities. The sample was then returned to Cutting Edge Ions for hydrogen implantation at a dose of $7 \times 10^{13}~\rm{cm}^{-2}$ and an energy of 9 keV. Implantation was followed by an anneal in 100~°C de-ionized water for 1 hour and a 3 minute rapid thermal at 420~°C.

\paragraph*{\TwoEightSi{} bulk sample} The hole burning on the bulk \TwoEightSi{} sample are performed on a sample of \TwoEightSi{} obtained from the Avogadro  project with 99.995\% \TwoEightSi{}, an oxygen concentration of less than $10^{14}$~cm$^{-3}$, and a natural carbon concentration of less than $5 \times 10^{14}$~cm$^{-3}$. The T centres were generated by irradiating the sample with 10 MeV electrons with an ionizing radiation dose of 320 kGy. After irradiation it was boiled in 100~$^\circ$C de-ionized water for 17 hours. This was followed by a sequence of anneals on a hotplate: 1 min at 300~°C, 30 min at 300~°C, 30 min at 350~°C, 60 min at 400~°C, and 50 min at 450~°C.


\paragraph*{Photonic chip cryogenics/sample mounting.} The SOI wafer is clamped to a copper plate with a plastic top hat to provide good thermal contact. The copper plate is mounted to an Attocube XYZ cryogenic positioner mounted on the still plate of a BlueFors dilution refrigerator with additional thermal lagging directly from the copper sample plate to the still plate. The copper sample plate has the option to house a 1/2 inch diameter, 1/8 inch thick, neodymium magnet with a room temperature magnetic field of $\sim$270 mT at the surface oriented vertically. The Bluefors refrigerator is operated at base temperature resulting in a still plate temperature of $1$--$1.2$~K, it may also be operated without circulating He-3 mixture resulting in a still plate temperature of $4.3$~K. The devices are aligned to polarization maintaining optical fibres (Thorlabs PS-PM980) arranged in a four-port fibre array fabricated by Fibertech Optica.

\paragraph*{Waveguide devices.} A strip waveguide with a width of $450$~nm is chosen for single-mode operation over the majority of T centre emission wavelengths. The total waveguide length is $360$~$\upmu$m long and each end terminates in a sub-wavelength grating coupler (GC) shown schematically in \cref{fig:device}(a). The GCs comprise a Bragg grating of silicon fins converging to a tapered waveguide section and diffract light of a chosen wavelength from the device plane into the fibre array or vice versa. A side-on view is shown in \cref{fig:device}(c). Each of the two GCs are designed for different wavelengths, one centred on the T centre ZPL wavelength and one red-shifted GC that covers a portion of the PSB spectrum. The two GC input/output ports are separated by $254$~$\upmu$m, twice the separation of fibres in the array, such that aligning the GCs to two fibre ports leaves an intermediate fibre port positioned over the waveguide centre.

\paragraph*{Waveguide photoluminescence.} A Thorlabs CPS450 continuous wave 450 nm laser is used to illuminate the middle of the waveguide device and the resulting emission is collected from each GC output separately and routed to a Princeton HRS-300 grating spectrometer with a choice of diffraction gratings providing resolutions in the range $0.10$--$0.002$~nm. The diffracted beam is imaged on an ultra-low noise Princeton NIRvana LN CCD camera.

\paragraph*{Waveguide photoluminescence excitation spectroscopy.} A Toptica CTL 1320 tunable laser is locked to a Bristol 871A-NIR wavemeter with a PID loop and used to scan across the wavelength range of interest. The resonant laser light is sent into the waveguide via the ZPL GC after a 1326 bandpass filter to remove non-resonant emission, a half-wave plate to align the polarization to the optimal GC polarization, and optional ND filters to set the power. A pick-off of the laser light is sent to a Thorlabs PDA20CS2 photodiode with a feedback to the laser to keep the power constant across the tuning range. Emitted light from the PSB GC is directed to a IDQ ID230 InGaAs/InP photon counter and later a IDQ ID281 superconducting nanowire single photon detector after remaining ZPL light is blocked by a 1340 nm longpass filter.

\paragraph*{Waveguide lifetime.} For above-band excitation a Picoquant LDH $965$~nm pulsed laser is used to excite the middle of the waveguide. For resonant excitation a tunable laser as described in the waveguide PLE section is pulsed with a Jenoptik electro-optic modulator (EOM) controlled with a Thorlabs MX10A EOM controller with re-biasing performed every 10 minutes. In both cases the emitted light from the PSB GC is collected in the same manner as in the waveguide PLE section and measured by an IDQ ID230 photon counter and the resulting single photon clicks are tagged with a Swabian Time Tagger Ultra. For the resonant excitation the measurement is repeated with the resonant laser detuned off the ZPL line to obtain a non-resonant decay curve, this is subtracted from the on-resonant decay curve to remove any fast emission from other species with the waveguide. The resulting luminescence decay curve is fit with an exponential decay.

\paragraph*{Waveguide hole burning.} Two-laser hole burning is done with a probe laser as described in the waveguide PLE section with an additional pump laser provided by either: (1) an equivalent Toptica CTL 1320 tunable laser for the at-field hole burning with pump power controlled by different ND filters (2) a Nanoplus 1326 DFB diode laser for the zero field hole burning. The Nanoplus laser uses a 1326 bandpass filter to clean up the emission along with a Thorlabs MX10A used for its power calibrated variable optical attenuator. In both cases the emitted light from the PSB GC is collected in the same manner as in the waveguide PLE section and measured with an IDQ ID230 photon counter for the at-field hole burning and a IDQ ID281 SNSPD for the zero-field hole burning.

\paragraph*{Waveguide $\mathbf{T_1}$.} Two tunable lasers as described in the waveguide PLE section are pulsed with two Brimrose AMR-100-1325 acousto-optic modulators (AOMs) with $>50$~dB extinction. Emission is collected as explained in the waveguide PLE section and measured with an IDQ ID281 SNSPD and the photon clicks are tracked with a Swabian Time Tagger Ultra. Pulsing of the AOMs and triggering of the time tagger was done with a SpinCore PulseBlaster USB TTL source.

\paragraph*{Bulk sample cryogenics.} Samples are mounted in a stress free manner to a sample rod lowered to the bottom of a Janis VariTemp liquid helium dewar. The samples are held at a temperature of $\sim1.4$~K with superfluid helium.

\paragraph*{Bulk sample photoluminescence.} A 200 mW 532 nm laser is used to excite the sample, the emission is collected and routed into a Bruker IFS125HR interferometer with a Ge diode detector. 

\paragraph*{Unpatterned SOI hole burning.} The sample is held in a reflective pocket in a liquid helium dewar. Two Toptica CTL 1320 tunable lasers are used as pump and probe with both lasers combined at a fibre beamsplitter. A pick off of the combined light is measured by a MenloSystems FPD610-FC-NIR fast detector (DC-600 MHz) and an Aglient DSO0954A oscilloscope operated as a frequency counter determines the pump-probe detuning by measuring the beat frequency. The probe light is modulated by an optical chopper before the fibre beamsplitter. The combined light is used to excite the sample and emission is collected and focused on to a Ge diode detector with a $1340$~nm longpass filter to remove ZPL laser light and a $1375\pm25$~nm bandpass filter to remove Raman excitation. The signal from the Ge diode is amplified with an Ithaco 1201 preamplifier and then a lock-in measurement is made with a Princeton Applied Research 5210 lock-in amplifier with the optical chopper reference frequency. A diagram of this experimental setup is shown in the Supplementary Information, \cref{fig:c13_hb}.

\paragraph*{\TwoEightSi{} hole burning.} To perform hole-burning spectroscopy in \TwoEightSi{} material, the sample is held in a reflective pocket in a liquid helium dewar and excited by pump and probe fields derived from a cavity-stabilized laser system: a Toptica DL100 Pro Pound-Drever-Hall locked to a Stable Laser Systems cavity using a Jenoptik PM1310 phase EOM. The stablized laser light (linewidth $<2.6$~kHz) is split between two paths separately modulated by ixBlue MPZ-LN-10 phase EOMs for stable frequency control. The two EOMs are driven by Stanford Research Systems SG386 signal generators. A single modulation sideband is mode-selected at the output of each modulator by custom made, temperature controlled etalons from Light Machinery (80~GHz FSR and $>100$ finesse). An additional, slow, modulation on the probe pathway is used for the subsequent lock-in measurement. Each path is separately amplified by Thorlabs CLD1015 optical amplifiers. The pump and probe are subsequently filtered with a 1326 clean-up filter and illuminate the sample. Luminescence from the sample is collected and focused onto a Ge diode detector with $1340$~nm longpass filter to remove the laser light and a $1375\pm25$~nm bandpass filter to remove light Raman scattered from the excitation lasers by the optical fibres. The electrical signal is amplified by an Ithaco 1201 preamplifier and a lock-in measurement is made with a Princeton Applied Research 5210 lock-in amplifier with the slow probe modulation as a reference. A diagram of this experimental setup is shown in the Supplementary Information, \cref{fig:sm_si28_hb_setup}.

\clearpage
\setcounter{figure}{0}
\renewcommand{\thefigure}{S\arabic{figure}}
\renewcommand{\thetable}{S\arabic{table}}

\title{Supplementary information: \\
    Waveguide-integrated silicon T centres}
\maketitle

\begin{figure}[h]
    \centering
    \includegraphics[width=17.2 cm]{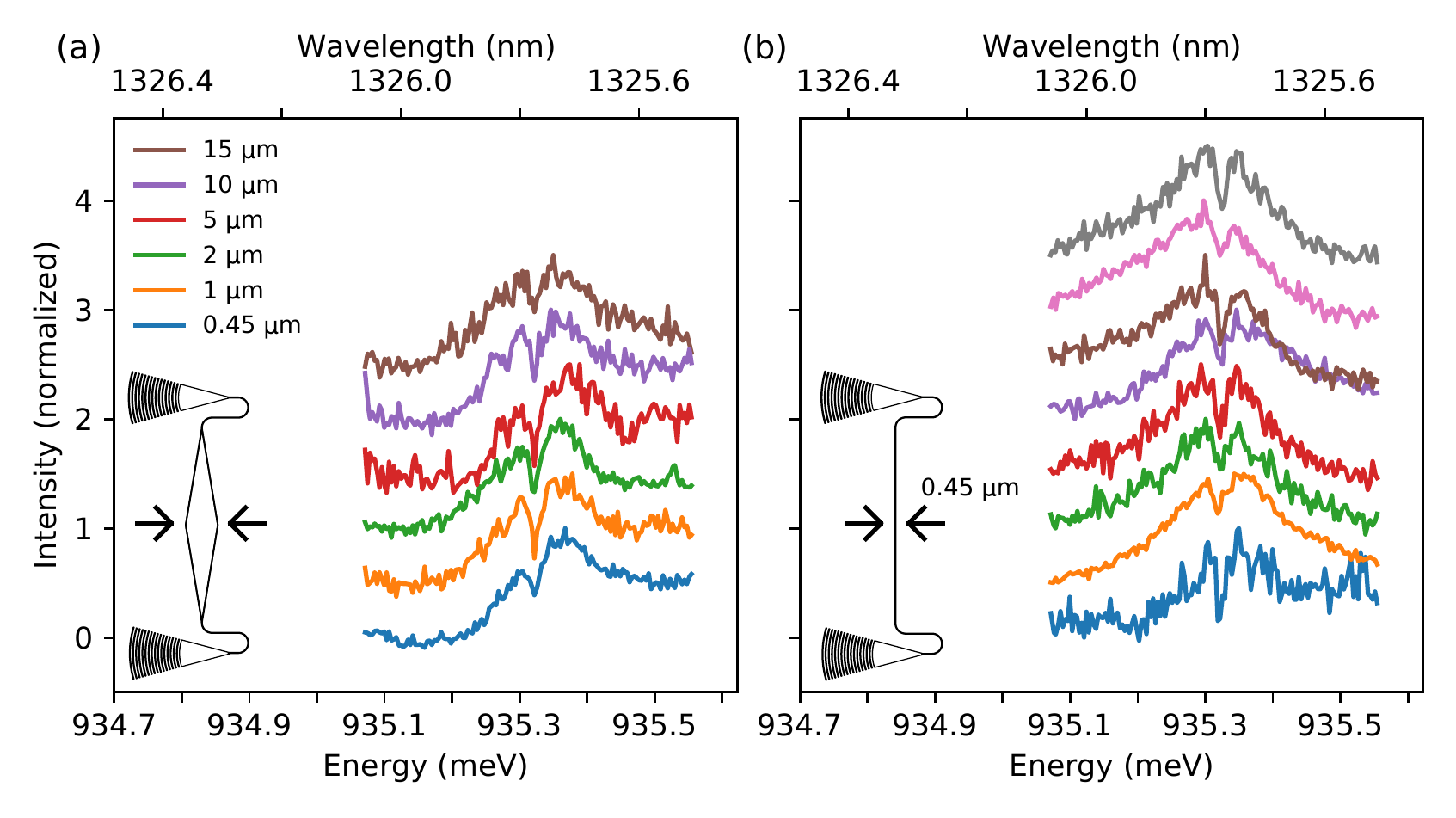}
    \caption{\textbf{Hole burning spectra in many devices.} Example spectra of nanophotonic waveguide devices with (a) different taper widths and (b) multiple copies of devices with no tapers, 0.45 um width. In both cases two laser hole burning is done to confirm T centre optical control. (Insets) Tapered and non-tapered devices, not to scale.}
    \label{fig:sm_waveguides}
\end{figure}

\textbf{Waveguide devices.} Upwards of 200 other GC coupled waveguide devices were fabricated in addition to the device presented in the main text. Some of these devices have a tapering waveguide that increases from the single-mode waveguide width ($0.45$~$\upmu$m) to a thicker multi-mode waveguide ($1$--$20$~$\upmu$m) and then back down again. An example $20$~$\upmu$m taper device is shown in the inset \cref{fig:sm_waveguides}(a). Two-laser hole burning PLE spectra of a range of taper device widths are shown in \cref{fig:sm_waveguides}(a), 
and hole-burning measurements of identical single-mode devices are shown in \cref{fig:sm_waveguides}(b). These results show the T centre recipe and patterning steps are compatible, and have a high device yield, essentially unity.

\begin{figure}
    \centering
    \includegraphics[width=13 cm]{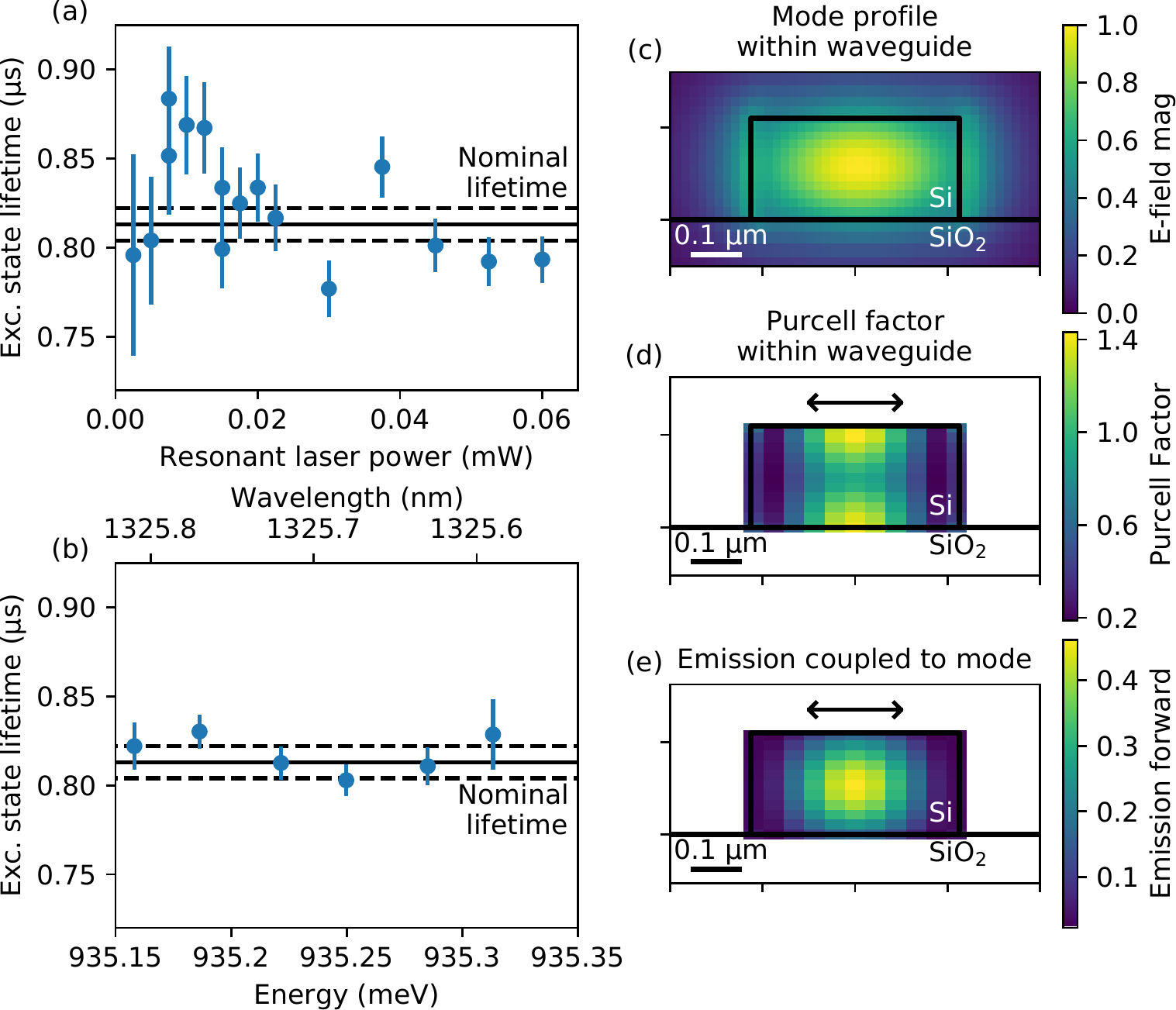}
    \caption{\textbf{Lifetime enhancement.} (a) Excited state lifetime versus resonant pump power showing no significant deviation from the nominal lifetime. (b) Excited state lifetime versus resonant pump wavelength showing no deviation from the nominal lifetime. All error bars are the single s.d. uncertainty. (c) Mode profile of the fundamental TE mode within a $0.45$~$\upmu$m wide waveguide at a wavelength of $1326$~nm. (d) Purcell factor for a dipole, oriented as shown by the arrow, situated at different locations within the same waveguide. (e) Coupling of emitted phonon sideband light from a dipole, oriented as shown by the arrow, to the fundamental TE mode.}
    \label{fig:sm_lifetime}
\end{figure}

\textbf{Lifetime enhancement.} As shown in the main text, we measure an ensemble-averaged TX$_0$ lifetime that varies slightly with excitation method. Resonant excitation through the waveguide mode results in 16$\%$ faster excited state decays than above-band excitation of the waveguide-coupled or bulk centres. We performed several additional measurements to identify the responsible mechanism. In \cref{fig:sm_lifetime}(a--b) we show the luminescence lifetime across a range of pump wavelengths and pump powers. In both cases the luminescence lifetime is unchanged to within the uncertainty of the measurement, from which we conclude that there is no measurable enhancement from stimulated emission.

We investigate potential Purcell enhancement effects by finite-difference time domain (FDTD) simulations of our single-mode waveguides in Lumerical. The results are plotted in \cref{fig:sm_lifetime}(c--e). The average Purcell enhancement (assuming T centres distributed evenly across the waveguide and dipoles aligned with the waveguide TE mode) is equal to the average of the Purcell distribution, \cref{fig:sm_lifetime}(d), weighted by the emission coupling profile, \cref{fig:sm_lifetime}(e). When the ensemble is resonantly excited through the waveguide at powers below saturation, an additional weighting by the excitation mode profile, \cref{fig:sm_lifetime}(c), is appropriate. This gives an average Purcell enhancement of $0.9$ and indicates that radiative emission should be slightly suppressed. An above bandgap excitation can be assumed to excite all T centres uniformly (rather than based on the mode profile) but this does not significantly change the expected enhancement.

\textbf{Unpatterned zero-field hole burning.} The sample is held in a reflective pocket in a liquid helium dewar. Two Toptica CTL 1320 tunable lasers are used as pump and probe with both lasers combined at a fibre beamsplitter. A pick off of the combined light is measured by a MenloSystems FPD610-FC-NIR fast detector (DC-600 MHz) and an Aglient DSO0954A oscilloscope operated as a frequency counter determines the pump-probe detuning by measuring the beat frequency. The probe light is modulated by an optical chopper before the fibre beamsplitter. The combined light is used to excite the sample and emission is collected and focused on to a Ge diode detector with a $1340$~nm longpass filter to remove ZPL laser light and a $1375\pm25$~nm bandpass filter to remove Raman excitation. The signal from the Ge diode is amplified with an Ithaco 1201 preamplifier and then a lock-in measurement is made with a Princeton Applied Research 5210 lock-in amplifier with the optical chopper reference frequency.. A diagram of this setup is shown in \cref{fig:c13_hb}.

\begin{figure}
    \centering
    \includegraphics[width=8.6 cm]{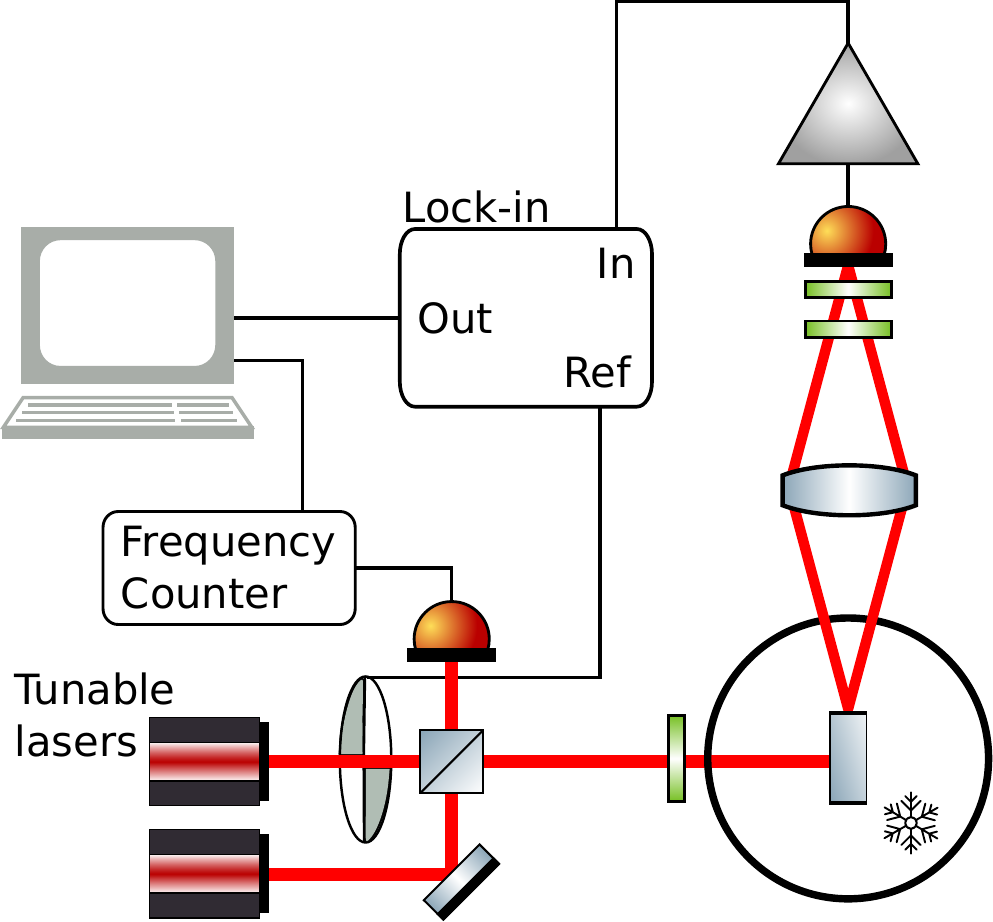}
    \caption{\textbf{Unpatterned zero-field hole burning.} The experimental setup used to measure hole burning in unpatterned SOI material. Pump and probe lasers excite the sample within a cryostat, a lock-in measurement is made on the collected sideband emission, and the pump-probe detuning is determined by a frequency counter measuring the beat frequency of the pump-probe interference on a fast detector.}
    \label{fig:c13_hb}
\end{figure}

\textbf{Hole burning in \TwoEightSi{}.}
To perform hole-burning spectroscopy in \TwoEightSi{} material, the sample is held in a reflective pocket in a liquid helium dewar and excited by pump and probe fields derived from a cavity-stabilized laser system: a Toptica DL100 Pro Pound-Drever-Hall locked to a Stable Laser Systems cavity using a Jenoptik PM1310 phase EOM. The stablized laser light (linewidth $<2.6$~kHz) is split between two paths separately modulated by ixBlue MPZ-LN-10 phase EOMs for stable frequency control. The two EOMs are driven by Stanford Research Systems SG386 signal generators. A single modulation sideband is mode-selected at the output of each modulator by custom made, temperature controlled etalons from Light Machinery (80~GHz FSR and $>100$ finesse). An additional, slow, modulation on the probe pathway is used for the subsequent lock-in measurement. Each path is separately amplified by Thorlabs CLD1015 optical amplifiers. The pump and probe are subsequently filtered with a 1326 clean-up filter and illuminate the sample. Luminescence from the sample is collected and focused onto a Ge diode detector with $1340$~nm longpass filter to remove the laser light and a $1375\pm25$~nm bandpass filter to remove light Raman scattered from the excitation lasers by the optical fibres. The electrical signal is amplified by an Ithaco 1201 preamplifier and a lock-in measurement is made with a Princeton Applied Research 5210 lock-in amplifier with the slow probe modulation as a reference. A diagram of this experimental setup is shown in \cref{fig:sm_si28_hb_setup}.

\begin{figure}
    \centering
    \includegraphics[width=17.2 cm]{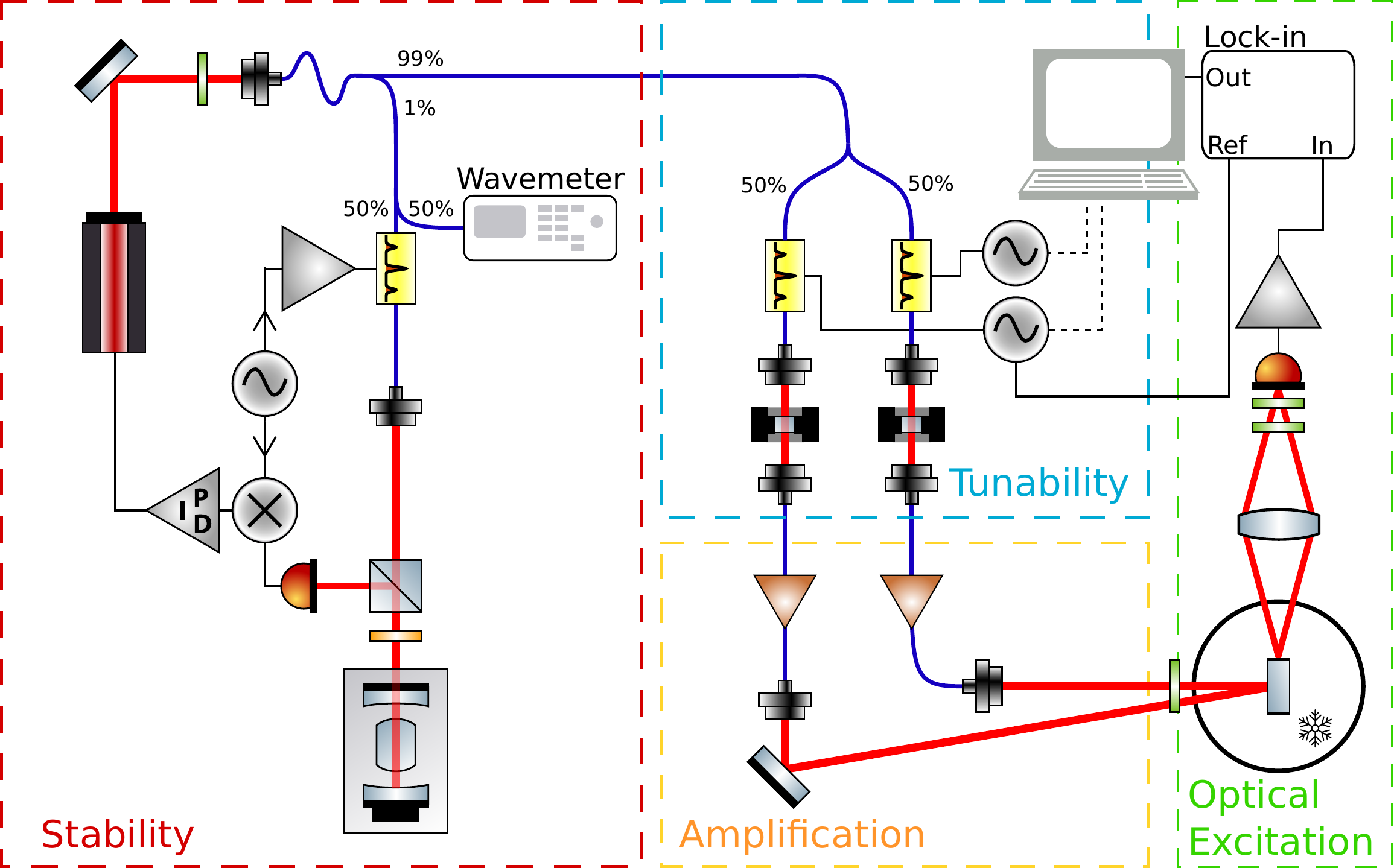}
    \caption{\textbf{Bulk \TwoEightSi{} hole-burning setup.} The experimental setup used to measure hole burning of the \TwoEightSi{} sample. Pump and probe light from a cavity-stabilized laser are individually modulated by phase EOMs and amplified before exciting the sample. Detuning of pump and probe is determined by the driving frequency of the two EOMs for which a mode-selecting etalon is used to isolate only one sideband. A lock-in measurement is made on the emitted sideband luminescence by slowly modulating the microwave drive of the probe EOM.}
    \label{fig:sm_si28_hb_setup}
\end{figure}

The large anti-holes in the \TwoEightSi{} hole-burning spectrum indicate that the instantaneous linewidth is less than or equal to the zero-field splitting and thus there is hyperpolarization occurring during the hole burning measurement. In order to determine the instantaneous linewidth we employ a three-level rate equation model.

At zero field we assume that the excited state is no longer split, but hyperfine coupling between the electron and hydrogen nuclear spins produces a zero-field splitting in the ground state. Hyperfine coupling to $^{13}$C in the sample can be ignored, the different isotopic T centre configurations are optically resolved in \TwoEightSi{} \cite{Bergeron2020}. The exact degeneracies of the ground state at zero field are unknown but the asymmetry in the results indicates it must be a singlet-triplet or triplet-singlet configuration---we test both with the rate equation model assuming that the triplet is degenerate.

A rate equation model is constructed for a three level system with arbitrary degeneracies, driven by two lasers with arbitrary power and a coupling determined by the detuning from a Lorentzian transition with arbitrary instantaneous homogeneous linewidth. As the measurement is slow compared to the excitation and decay rates we solve this system of equations in equilibrium ($\dot{N}_i=0$) and take the excited state population to be proportional to the measured signal. It was assumed in the model that no population is transferred directly between ground states as $T_1>16$~s~\cite{Bergeron2020} is much longer than the measurement times.

\begin{figure}
    \centering
    \includegraphics[width=8.6 cm]{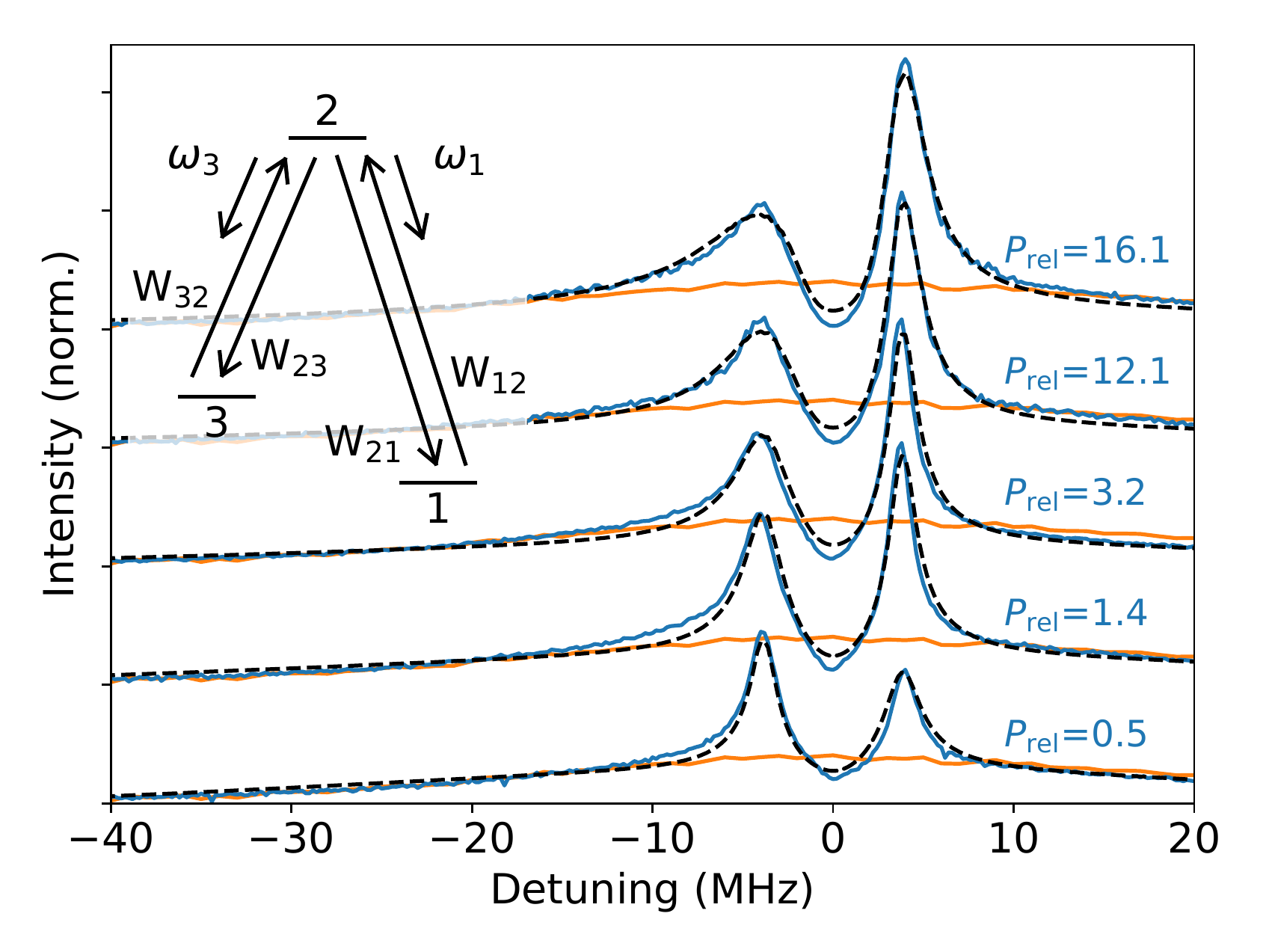}
    \caption{\textbf{\TwoEightSi{} hole burning.} Hole burning of the \TwoEightSi{} sample across a range of pump powers denoted by the ratio $P_{\rm rel}=P_{\rm pump}/P_{\rm probe}$ with a fixed probe power of $62$~$\upmu$W. A single laser scan is shown in orange. The intensities are normalized to the value at $-20$~MHz, far from the effects of the pump. The dashed line is the simultaneous fit of all data with the rate equation model described in the text. (Inset) The three level model with all rates considered shown schematically. For the presented fit, state 3 is a triply degenerate level.}
    \label{fig:sm_si28_hb}
\end{figure}

The steady-state rate equation is:
\begin{gather}
\label{eq:rate_first}
 \begin{bmatrix} \dot{N}_{1} \\ \dot{N}_{2} \\ \dot{N}_{3} \end{bmatrix}
 = 
  \begin{bmatrix} 0 \\ 0\\ 0 \end{bmatrix}
  =
  \begin{bmatrix}
   -W_{12} & W_{21} + \omega_1 & 0 \\
   W_{12} & -W_{21} - W_{23} - \omega_1 - \omega_3 & W_{32} \\
   0 & W_{23} + \omega_3 & -W_{32} \\
   \end{bmatrix}
   \begin{bmatrix} N_{1} \\ N_{2} \\ N_{3} \end{bmatrix}
\end{gather}
Where the rates are shown schematically in \cref{fig:sm_si28_hb}. Requiring that the system is closed ($N_1+N_2+N_3=1$) gives the excited state population
\begin{gather}
  N_{2}^{\rm SS} = \frac{W_{12}W_{32}}{W_{12}W_{32} + W_{12}(W_{23} + \omega_3) + W_{32}(W_{21} + \omega_1)}.
  \end{gather}
 The relationship between upward and downward rates is determined by the relative degeneracies, $n_{i}$, as~\cite{Siegman1986}
\begin{gather}
W_{ij} = \frac{n_j}{n_i}W_{ji}.
\end{gather}
The degeneracies also influence the relative spontaneous decay rates as
\begin{gather}
\omega_i = \frac{n_i}{n_1 + n_3}\omega,
\end{gather}
where $\omega$ is the experimentally observed decay rate, $\omega=0.94\pm0.01$~$\upmu$s.

The driving rates $W_{i2}$ are the peak pump/probe driving rates, $W_{\rm pump}$ / $W_{\rm probe}$, multiplied by a normalized Lorentzian prefactor, $L(f, \Gamma)$, with linewidth $\Delta\omega_{\rm hom}$ and a relative frequency given by the pump/probe detuning, $f_{\rm pump/probe}$; the ground state splitting, $\pm \Delta_{\rm g}/2$; and the detuning of the sub-ensemble within the inhomogeneous linewidth, $f_{\rm inhom}$, resulting in
\begin{gather}
\begin{split}
W_{i2}(f_{\rm pump}, f_{\rm probe}, f_{\rm inhom}) &= L(f_{\rm pump} - f_{\rm inhom} \mp \Delta_{\rm g}/2, \Delta\omega_{\rm hom}) W_{\rm pump} \\
& + L(f_{\rm probe} - f_{\rm inhom} \mp \Delta_{\rm g}/2, \Delta\omega_{\rm hom}) W_{\rm probe},
\end{split}
\end{gather}
where $i=1$ corresponds to the `-' case and $i=3$ corresponds to the `+' case. $W_{\rm pump}$ and $W_{\rm probe}$ are the powers of the pump and probe laser respectively. These values can be related to the laser powers by comparing the saturation power to the saturation rate $W_{\rm sat} = 1./(2\tau_{\rm exc})$, where $\tau_{\rm exc}$ is the excited state lifetime~\cite{Siegman1986} and the saturation power must be measured experimentally. 

The final signal as a function of pump and probe frequencies is the excited state population at steady state, $N_{2}^{\rm SS}$, integrated over all sub-ensembles within the inhomogeneous line, accounted for by integrating over all frequencies with a normalized Gaussian-Lorentzian sum distribution, $I(\Delta f, m, \Gamma)$, as a prefactor with the inhomogeneous linewidth, $\Gamma_{\rm inhom}$, and a free parameter, $m$, for the Gaussian-Lorentzian proportion. The signal, $S(f_{\rm pump}, f_{\rm probe})$, as a function of the pump and probe frequencies is
\begin{gather}
\label{eq:rate_last}
S(f_{\rm pump}, f_{\rm probe}) = \int_{-\infty}^{\infty} I(f_{\rm inhom}, m, \Gamma_{\rm inhom.}) N_{2}^{\rm SS}(f_{\rm pump}, f_{\rm probe}, f_{\rm inhom}) df_{\rm inhom}.
\end{gather}

We performed a simultaneous fit of five different scans with five different pump-probe ratios. Eq.~\ref{eq:rate_last} is integrated numerically and the data is fit with an arbitrary offset and scaling factor. The fit was repeated for the two possible singlet-triplet configurations, ($n_1=3$, $n_2=4$, $n_3=1$) and ($n_1=1$, $n_2=4$, $n_3=3$),  of which only the latter matches the data, indicating that the triplet is higher energy (as shown in the inset of \cref{fig:sm_si28_hb}). The results give a homogeneous linewidth of $0.69\pm0.02$~MHz and a zero-field splitting of $3.848\pm0.007$~MHz. The fits are shown in \cref{fig:sm_si28_hb}.

\begin{figure}[h]
    \centering
    \includegraphics[width=8.6 cm]{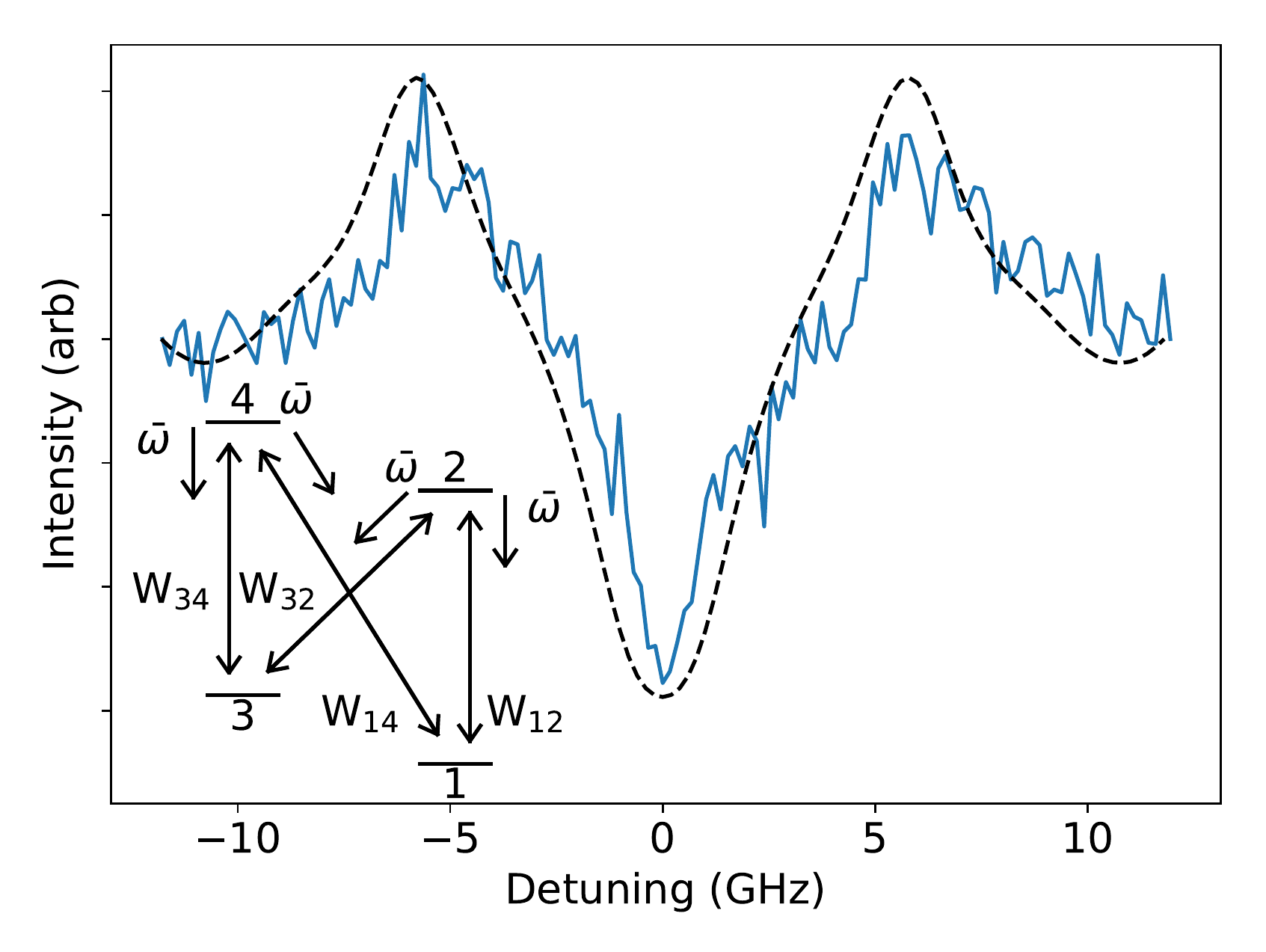}
    \caption{\textbf{At-field hole burning zero-parameter fit.} An at-field hole burning spectrum of $^{\rm nat}$Si waveguide material along with a zero parameter fit using a four level rate equation. (Inset) Schematic illustration of the rates involved in the four level rate equation.}
    \label{fig:4_level_rate}
\end{figure}

\textbf{At-field hole burning spectrum.} The at-field hole burning spectrum can be modeled using a four-level rate equation formulated in the same manner as the three-level rate equation model 
\begin{gather}
\label{eq:rate_first_4}
 \begin{bmatrix} \dot{N}_{1} \\ \dot{N}_{2} \\ \dot{N}_{3} \\ \dot{N}_{4}\end{bmatrix}
  =
   \begin{bmatrix} 0 \\ 0 \\ 0 \\ 0\end{bmatrix}
   =
  \begin{bmatrix}
   -W_{12}-W_{14} & W_{21} + \omega_{21} & 0 & W_{41} + \omega_{41} \\
   W_{12} & -W_{21} - W_{23} - \omega_{21} - \omega_{23} & W_{32} & 0\\
   0 & W_{23} + \omega_{23} & -W_{32} - W_{34} & W_{43} + \omega_{43}\\
   W_{14} & 0 & W_{34} & -W_{41} - W_{43} - \omega_{41} - \omega_{31}\\
   \end{bmatrix}
   \begin{bmatrix} N_{1} \\ N_{2} \\ N_{3} \\ N_{4}\end{bmatrix}
\end{gather}
Where these rates are shown schematically in \cref{fig:4_level_rate} and we assume all degeneracies are equal. Thus $W_{ij} = W_{ji}$ and likewise all decay rates $\omega_{ij}$ are equal to $\bar{\omega} \equiv 1./(2\tau_{\rm exc})$. Once again $T_1$ is long and we can neglect spin flips in the ground state.

We solve this system in the steady state and take the sum of population in the excited states 2 and 4, $N^{\rm SS}_{2+4}$, to be proportional to the measured signal. The excited state population in steady state is
\begin{equation}
    N^{SS}_{2+4} = \frac{\splitfrac{W_{12}W_{14}W_{32} + W_{12}W_{14}W_{34} + W_{12}W_{32}W_{34} + W_{12}W_{34}\bar{\omega}}{ + W_{14}W_{32}W_{34} + W_{14}W_{32}\bar{\omega} + W_{12}W_{32}\bar{\omega} + W_{14}W_{34}\bar{\omega}}}{\splitfrac{2W_{12}W_{14}W_{32} + 2W_{12}W_{14}W_{34} + W_{12}W_{14}\bar{\omega} + 2W_{12}W_{32}W_{34} + 3W_{12}W_{32}\bar{\omega} + 2W_{12}W_{34}\bar{\omega} + W_{12}\bar{\omega}^2}{+2W_{14}W_{32}W_{34} + 2W_{14}W_{32}\bar{\omega} + 3W_{14}W_{34}\bar{\omega} + W_{14}\bar{\omega}^2 + W_{32}W_{34}\bar{\omega} + W_{32}\bar{\omega}^2 + W_{34}\bar{\omega}^2}} \,.
\end{equation}
The driving rates for the four different transitions are
\begin{gather}
\begin{split}
W_{ij}(f_{\rm pump}, f_{\rm probe}, f_{\rm inhom}) &= L(f_{\rm pump} - f_{\rm inhom} \mp \Delta_{\rm g}/2 \mp \Delta_{\rm e}/2, \Delta\omega_{\rm hom}) W_{\rm pump} \\
& + L(f_{\rm probe} - f_{\rm inhom} \mp \Delta_{\rm g}/2 \mp \Delta_{\rm e}/2, \Delta\omega_{\rm hom}) W_{\rm probe} \,,
\end{split}
\end{gather}
where the excited state splitting is $\Delta_{\rm e} = g_h\mu_BB_0$ and the ground state splitting is $\Delta_{\rm g} = g_e\mu_BB_0$ with hole and electron Land\'{e} g-factors, $g_{h,e}$, the Bohr magneton, $\mu_B$, and the static magnetic field, $B_0$.

As before, the final signal as a function of pump and probe frequencies is the integral of $N^{SS}_{2+4}$ over the inhomogeneous frequency shift weighted by a Gaussian-Lorentzian sum prefactor
\begin{gather}
S(f_{\rm pump}, f_{\rm probe}) = \int_{-\infty}^{\infty} I(f_{\rm inhom}, m, \Gamma_{\rm inhom.}) N^{\rm SS}_{2+4}(f_{\rm pump}, f_{\rm probe}, f_{\rm inhom}) \,df_{\rm inhom} \,.
\end{gather}

With this four-level rate equation we perform a zero-parameter fit to the at-field hole burning data. We assume the hole to anti-holes splitting, 6~GHz, is from the ground state splitting, $\Delta_{\rm g}$ = 6~GHz, indicating a 210~mT magnetic field. We average together contributions from the twelve different orientations that have excited state splittings determined by the two hole g-factors for a field along the $\langle100\rangle$ axis, $g_h = \{0.91, 2.55\}$ with degeneracies 4 and 8 respectively~\cite{MacQuarrie2021}. These give excited state splittings of $\Delta_{\rm e}=\{2.73, 7.65\}$~GHz. Further parameters used are: $\Delta\omega_{\rm hom} = 600$~MHz from the fit low-power linewidth, $W_{\rm pump} = 10~W_{\rm sat}$ and $W_{\rm probe}=0.3~W_{\rm sat}$ from the saturation power fit to the hole burning linewidth versus pump power. The resulting simulated spectrum, after normalizing in the same manner as the data, is shown in Fig.~\ref{fig:4_level_rate}. The close fit strongly suggests that we've fairly attributed the anti-holes to the ground state splitting and that the hole g-factors predominantly match the expected set. Furthermore, the magnetic field value of 210~mT is consistent with the magnetic field measured at room temperature, $270$~mT, after accounting for the $\sim14\%$ reduction from the neodymium spin reorientation transition below 135 K~\cite{Tokuhara1985}.

\begin{figure}[h]
    \centering
    \includegraphics[width=8.6 cm]{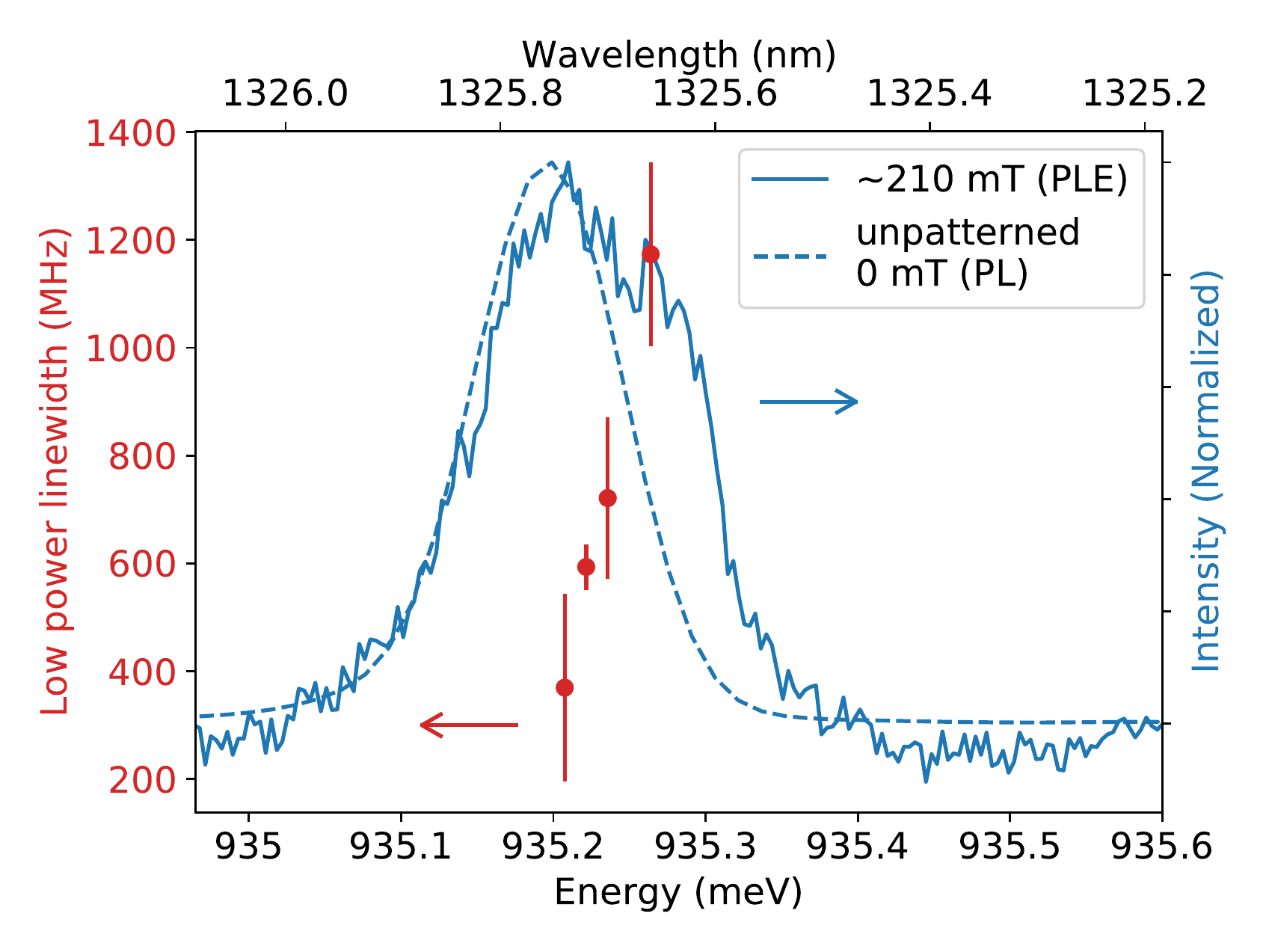}
    \caption{\textbf{At-field hole burning versus wavelength.} (Red) fit low power limit linewidths from hole burning at different pump wavelengths. (Blue) One laser PLE of the waveguide device under test, and PL of unpatterned material.}
    \label{fig:sm_hb_vs_wl}
\end{figure}

\textbf{At-field hole burning at different wavelengths.} In addition to the hole burning at field presented in the main text, hole linewidth versus power was measured at additional detunings on the ZPL line. The low-power hole linewidth at different laser wavelengths, Fig.~\ref{fig:sm_hb_vs_wl}, shows a trend of homogeneous linewidth versus wavelength that may be due to strain within the sample from mounting. This suggests that strain (and equivalently electric fields) can make the T centre more/less susceptible to spectral diffusion.

\textbf{Interference visibility.} From the homogeneous linewidths we can determine the expected emission interference visibility (or photon indistinguishability). Using the formalism from Ref.~\cite{Kambs2018} the visibility, $V$, depends only on the instantaneous homogeneous linewidth, ie, the pure dephasing linewidth ($\Gamma_{\rm PD}$ in Ref.~\cite{Kambs2018}); the emitter lifetime; and slow spectral diffusion. With appropriate feedback slow spectral diffusion can be made negligible and the visibility only depends on linewidth and lifetime. In Tab.~\ref{tab:visibilities} we tabulate expected interference visibilities for the samples measured in this manuscript. 

Additionally, we tabulate the cavity properties required to achieve an interference visibility of $56\%$. This is the photon indistinguishability required to entangle two spin-photon emitters by the Barrett-Kok protocol \cite{Barrett2005} beyond the Bell violation entanglement threshold \cite{Hensen2015}. We determine the necessary Purcell factor, $F_P$, accounting for the phonon sideband and assuming unit radiative efficiency. Under these assumptions the Purcell enhanced lifetime, $\tau_{\rm cav}$,is given by~\cite{Faraon2012}
\begin{equation}
    1/\tau_{\rm cav} = (F_P+1)/\tau_{\rm zpl} + 1/\tau_{\rm psb}.
\end{equation}
$\tau_{\rm zpl}$ and $\tau_{\rm psb}$ are the ZPL and PSB emission lifetimes given by $\tau_{\rm exc}/\eta_{\rm zpl}$ and $\tau_{\rm exc}/(1 - \eta_{\rm zpl})$ respectively where $\eta_{\rm zpl}$ is the Debye-Waller factor, 0.23~\cite{Bergeron2020}. We determine the cavity Q-factor required to achieve such a Purcell enhancement in a nanophotonic cavity with a modal volume of $\frac{1}{2}(\lambda/n)^3$.
\renewcommand{\arraystretch}{1.2} 
\begin{table}[h]
\centering
\begin{tabular}{lllll}
 & $\Delta\omega_{\rm hom}$  (MHz) & $F_{P}$ & $Q$ & $V$ \\
 \hline
 \hline
 \multirow{4}{*}{$^{\rm nat}$Si waveguides} &  \multirow{2}{*}{67 (Measured)}& \multicolumn{2}{c}{No cavity}  & 0.003 \\
  & & 2200 & 14,000 & 0.56 \\
 & \multirow{2}{*}{11 (Low-power limit)} & \multicolumn{2}{c}{No cavity} & 0.015 \\
 & & 350  & 2300 & 0.56\\
  \hline
 \hline
 \multirow{2}{*}{$^{\rm 28}$Si }&  \multirow{2}{*}{0.69 (Measured)} &  \multicolumn{2}{c}{No cavity} & 0.25\\
& &  13 & 85 & 0.56 \\
\end{tabular}
\caption{\textbf{Interference visibility.} Interference visibility ($V$) of emission from T centres in different environments according to their measured or inferred instantaneous homogeneous linewidths ($\Delta\omega_{\rm hom}$). Purcell factors ($F_{P}$) required to reach 56\% interference visibility and the corresponding cavity quality factors ($Q$) are included in each case.}
\label{tab:visibilities}
\end{table}
\end{document}